\documentclass[10pt,twocolumn]{article}

\usepackage[utf8]{inputenc}
\usepackage[T1]{fontenc}
\usepackage[margin=1in]{geometry}
\usepackage{amsmath,amssymb}
\usepackage{booktabs}
\usepackage{longtable}
\usepackage{graphicx}
\graphicspath{{figures/}}
\usepackage{rotating}   
\usepackage{adjustbox}  
\usepackage{subcaption}
\usepackage{xcolor}
\usepackage[round]{natbib}
\usepackage[hidelinks]{hyperref}
\usepackage{microtype}
\newcommand{\WE}{\mathbf{W}_{E}}      
\newcommand{\WU}{\mathbf{W}_{U}}      
\newcommand{\Eindex}{\mathcal{E}}     
\newcommand{\bpb}{\mathrm{bpb}}

\title{\textbf{Reading Without a Reader:\\
Large Language Models Collapse the Brain's Dissociated\\
Reading and Writing Systems into a Single Entangled Code}}

\author{Diego Saldaña Ulloa\thanks{diego.ulloa13@hotmail.com}}

\begin{document}
\maketitle

\begin{abstract}
In the literate human brain, reading and writing doubly dissociate: a ventral
decoding route (impaired in pure alexia) and a fronto-parietal encoding route (pure
agraphia), sharing a partial orthographic core. A decoder-only large language model
(LLM) instead drives both from one autoregressive path optimized on text (a
\emph{cultural} invention, not an evolved instinct). We ask how entangled that
mechanism is, comparing an input-side ``reading code'' $\WE$ with an output-side
``writing code'' $\WU$ via an index $\Eindex\in[0,1]$ (CKA, Procrustes residual,
mutual $k$-NN) calibrated against an independent-init floor and a tied ceiling. On
GPT-2, OPT and Pythia (14M--1.4B), untied models hold one
\emph{coupled but sub-ceiling} code ($\Eindex=0.23$--$0.35$, far above floor) on a
non-monotonic couple-then-differentiate trajectory, $\WU$ drifting
$\sim$3.2$\times$ farther than $\WE$ in every frequency decile. We then report an equally
informative negative: the matching behavioural test (that comprehension and production fail together rather than dissociate) cannot be run. For minimal pairs the analogue of pure alexia is empty by theorem, since greedy
production implies a vocabulary-wide argmax and so wins the pairwise ranking.
Differential-damage indices under ablation are not scale-identified: heavy-tailed
damage makes linear standardizations collapse onto their larger term, and the rank
transform that fixes this is bounded, so its null saturates. And both scores contain the
target's log-probability, which alone explains most of their variance and
manufactures the coupling it seems to reveal. We withdraw the coupling statistic, the
cross-level bridge and an encoder/decoder separation measure. In a
model reading and writing off one next-token distribution, no pair of output-side
scores isolates either ability (entanglement that needs no index to see).
By analogy, not homology, this situates LLMs in the space of possible minds.


\end{abstract}

\section{Introduction}

Spoken language is, on the dominant view, an evolved biological instinct: it
emerges spontaneously in every typically developing child, in every human
culture, without explicit instruction \citep{pinker1994}. Reading and writing
are not. Written language is a recent cultural invention (the earliest scripts
are roughly five millennia old), far too young to have been shaped by natural
selection. Literacy is instead grafted onto cortex that evolved for other
purposes, a process Dehaene and colleagues call \emph{neuronal recycling}: the
acquisition of reading repurposes a region of the ventral occipitotemporal
cortex, the visual word form area (VWFA), that becomes selective for letter
strings only after a reader learns to read \citep{dehaene2007recycling, dehaene2011vwfa}. The cultural, acquired character of literacy is
visible even in its cognitive byproducts: explicit awareness of speech as a
sequence of phonemes does not arise spontaneously but is a \emph{consequence} of
alphabetic literacy (illiterate adults fail the phoneme-manipulation tasks that
late-literates pass; \citealp{morais1979}), and the cortical reorganization induced
by reading acquisition leaves a measurable anatomical signature
\citep{carreiras2009}.

Crucially, in the human brain the cultural skill of literacy is realized not as
one faculty but as \emph{two}. Reading (decoding visual symbols toward meaning)
and writing (encoding meaning toward orthographic and motor output) doubly
dissociate: a focal lesion can abolish reading while sparing writing (pure
alexia without agraphia, the classical disconnection syndrome described by
\citet{dejerine1892}), and another can abolish writing while sparing reading
(pure agraphia, implicating a distinct graphemic and fronto-parietal encoding
system; \citealp{caramazza1987buffer, roux2009exner}).
The two systems are partly separable yet share a common orthographic core: the
same ventral occipitotemporal and left inferior frontal populations are engaged
during both reading and spelling \citep{purcell2017shared}, and even decoding
alone is itself multi-route \citep{coltheart2001drc}. The human reading/writing
architecture is thus best described as two dissociable systems over a shared
orthographic substrate.

Large language models acquire functional literacy under radically different
conditions. A decoder-only transformer has no pre-existing evolved substrate to
recycle; it optimizes the cultural artifact of written text directly, using a
single general-purpose mechanism (self-attention; \citealp{vaswani2017}) whose
in-context behavior is carried by identifiable induction circuits
\citep{olsson2022, elhage2021framework}. Most importantly, the architecture
\emph{collapses} reading and writing into one entangled autoregressive pass.
There is no anatomical separation between a decoding network and an encoding
network; there is one stack of weights that both consumes and produces text. The
nearest structural analog to the brain's read/write split is the pair formed by
the input embedding $\WE$, which maps tokens into the residual stream (the
``reading code''), and the output unembedding $\WU$, which maps the residual
stream back to token logits (the ``writing code''). In \emph{tied} models
(GPT-2, OPT) these are forced to be identical by construction
\citep{presswolf2017, inan2017tying, radford2019, zhang2022opt}; in
\emph{untied} models (the Pythia / GPT-NeoX suite) they are free to be learned
separately \citep{biderman2023}. We stress at the outset that this is an
\emph{analogy, not a homology}: $\WE$ and $\WU$ are input- and output-side
token-coding matrices in the same modality and processing stage, not literal
reading and writing networks, and the alexia/agraphia dissociation is a
motivating contrast rather than a measured correspondence. (We also note that the
residual-stream framework of \citet{elhage2021framework} uses ``write/read'' for
the opposite direction: components \emph{write to} and \emph{read from} the
stream; our labels instead track the brain's input-reading/output-writing
convention.) We make no claim about ``understanding.''

This framing raises a precise empirical question. If the brain realizes literacy
as two dissociable systems over a shared core, and a decoder-only LLM realizes
it as one statistical code, then we may read LLMs as occupying a \emph{distinct
point in the space of possible minds} (achieving functional literacy by a route
the human brain does not take) rather than as a model of the human
reading/writing brain \citep{sloman1984, shanahan2025, mitchellkrakauer2023}. We
are deliberate about what is and is not at stake here: because a decoder-only
architecture commits to a single forward path, \emph{some} read/write coupling is
expected a priori, so the existence of coupling is not itself a discovery. Our
empirical contribution is to \emph{quantify} how far that coupling sits from a
genuine-dissociation floor and a forced-collapse ceiling, and to characterize its
training trajectory. The ``distinct point in the space of minds'' reading is offered
as an interpretive lens on those measurements, not as a proven fact. Geometrically,
we define an entanglement index $\Eindex \in [0,1]$ that fuses three established
representation-similarity measures (linear CKA, \citealp{kornblith2019cka};
orthogonal-Procrustes residual with per-token cosine, \citealp{smith2017offline};
and mutual $k$-NN overlap, \citealp{huh2024platonic}) between $\WE$ and $\WU$,
calibrated against an independent-initialization floor, a shuffled-token null,
and the tied ceiling.

We also set out to test the matching behavioural prediction, contrasting
comprehension (log-probability discrimination) with production (greedy generation)
and asking whether the two dissociate as they do in the brain. That test, and two
redesigns of it, could not be made valid, and we report why as a result in its own
right: comprehension and production, as output-side scores define them, are both
functionals of a single next-token distribution, so any two scalar summaries of it
share terms, and those shared terms generate the very coupling one is trying to
detect.
The reading and writing of these models are not merely entangled in the weights;
at the output they are not separately addressable. We situate all of this against
the wider debate on formal versus functional linguistic competence
\citep{mahowald2024, fedorenko2024language} and symbol grounding
\citep{harnad1990symbol}.

\paragraph{Contributions.}
\begin{itemize}
  \item \textbf{A consolidating replication.} Six of our ten probes reproduce
        previously established findings (induction heads, \citealp{olsson2022}; the
        literacy data-efficiency gap; form-before-meaning, \citealp{mahowald2024};
        compression scaling; verbatim memorization, \citealp{carlini2023quantifying};
        and an imitation-vs-innovation gap, \citealp{yiu2024imitation}), run
        uniformly on one Pythia suite on a single GPU. We claim these as
        confirmation, not discovery; our new contributions are the items below.
  \item \textbf{A brain-anchored read/write axis for interpretability.} We
        reframe the input embedding and output unembedding of a decoder-only LLM
        as a ``reading code'' $\WE$ and a ``writing code'' $\WU$, motivated by
        the brain's doubly dissociable reading and writing systems, and
        introduce a controlled entanglement index $\Eindex$ with an
        independent-init floor, a shuffled-token null, and a tied ceiling.
  \item \textbf{One coupled, sub-ceiling read/write code in the weights.}
        In untied Pythia models $\Eindex$ sits far above the independent-init
        floor but well below the tied ceiling: one coupled-but-sub-ceiling code
        rather than either two independent codes or one shared code. The trajectory is non-monotonic, a
        ``couple-then-differentiate'' program (peak $\mathrm{CKA}=0.494$ at
        step~4000), and the differentiation is asymmetric (in pythia-160m $\WU$
        moves $\sim\!3.2\times$ farther than $\WE$ in every frequency decile), so
        the writing code, not the reading code, does the moving.
  \item \textbf{Three reasons the behavioural test cannot be run, which are
        themselves a result.} We attempted the matching behavioural prediction (that comprehension and production should fail together rather than dissociate) and neither it nor two redesigns could be made valid
        (Section~\ref{sec:res-limits}). For minimal pairs the model analogue of
        pure alexia is empty by theorem, because greedy production implies a
        vocabulary-wide argmax and hence wins the pairwise ranking. Indices of
        differential damage under ablation are not scale-identified: the damage
        distributions are heavy-tailed enough that every linear standardization
        collapses the index onto its larger term, and the rank transform that fixes
        this exactly is bounded, so its family-wise null saturates. And the two
        scores share the target's log-probability, which alone accounts for most of
        the variance in both and manufactures the coupling it appears to reveal.
        We report the withdrawn claims explicitly and keep the diagnosis, which is
        the sharper statement: in a model whose reading and writing are read off
        one next-token distribution, no pair of output-side scores isolates one
        ability from the other.
  \item \textbf{An honest accounting of nulls.} We report prominently that we find
        \emph{no evidence} for the ``compression decoupled from reasoning''
        sub-claim (the point estimate runs opposite, Pearson $-0.68$, $p=0.14$,
        $n=6$, underpowered) and withdraw it, that ablating the induction circuit
        leaves the orthographic word-form gap intact (fraction remaining $1.009$),
        and that the scale trend in $\Eindex$ is flat and underpowered. We also
        withdraw, with their reasons, a behavioural coupling statistic and its sign
        test, a cross-level geometry$\rightarrow$behaviour bridge that was computed
        against it, and a representational encoder/decoder separation measure.
  \item \textbf{A position in the space of possible minds (as a lens).} Granting
        that single-path coupling is architecturally expected, what the
        quantification adds is its calibration (coupled, but well short of the tied ceiling, and reached by a non-monotonic route) alongside the
        finding that the distinction resists output-side operationalization at all.
        Together these invite reading decoder-only LLMs as realizing functional
        literacy through a single entangled statistical code: a distinct point in
        the space of minds, offered as an interpretive framing and an analogy, not
        a homology, to the human reading/writing brain.
\end{itemize}

\paragraph{Relation to prior work.}
Our geometric result is closest to \citet{lopardo2026weighttying}, who first
reported that untied token embeddings drift toward the output space, with the
output-side matrix moving more than the input-side one. We credit that work for
the asymmetric-drift trajectory; our Experiment~7 is a controlled,
properly-nulled replication of it on the read/write axis (the floor, null,
ceiling, and frequency-decile controls are detailed in
Section~\ref{sec:related-transformers} and Section~\ref{sec:methods-eindex}). Our
conceptual frame extends \citet{mahowald2024}, who dissociate \emph{formal} from
\emph{functional} linguistic competence: we add a second, orthogonal
\emph{read/write} axis grounded in the neuropsychology of literacy and carry it
down to weight geometry. On the behavioral side, comprehension/production
asymmetries in LLMs have been studied directly \citep{lam2025prodinterp,
pickering2013integrated}; we differ by attempting to bridge that behavioral
dissociation to the underlying $\Eindex$ weight geometry, and by reporting that
the bridge cannot be built (not because the correlation is weak, but because the
behavioural term does not survive scrutiny as a measurement;
Section~\ref{sec:res-limits}). Finally, where \citet{shanahan2025} and
\citet{mitchellkrakauer2023} argue on conceptual grounds that LLMs are a different
kind of system, our contribution is to supply controlled, mechanistic evidence at
the level of weights, with its nulls reported plainly, for one specific way in
which they differ from the human reading/writing brain.

\section{Background and Related Work}
\label{sec:related}

Our argument draws together four bodies of work: (i) the cognitive neuroscience of
reading and writing, which establishes that human literacy is realized by two
partially dissociable cortical systems; (ii) the cultural-artifact view of written
language and the resulting human--machine gap in sample efficiency; (iii) the
interpretability of transformers and the geometry of their input/output embeddings,
which gives a tractable weight-level analog of a ``reading code'' and a ``writing
code''; and (iv) the debate over whether large language models (LLMs) are best
understood as a distinct \emph{kind} of mind.

\subsection{Reading and writing in the brain}
\label{sec:related-brain}

Reading is not an evolved faculty. On the \emph{neuronal recycling} hypothesis
\citep{dehaene2007recycling}, the cultural invention of writing (too recent for natural selection to have shaped a dedicated organ) colonizes and repurposes
phylogenetically older visual and language cortex, inheriting its computational
constraints. The empirical anchor is the Visual Word Form Area, a patch of left
ventral occipitotemporal cortex that becomes selectively responsive to orthographic
strings only in literate readers \citep{dehaene2011vwfa}. Learning to read measurably
reorganizes the brain, sharpening visual responses and reshaping functional and even
structural connectivity \citep{carreiras2009}. Reading is thus an acquired overlay on
evolved hardware, a point we return to when interpreting the \emph{acquisition} of
orthographic structure in a from-scratch transformer.

Crucially for our thesis, reading (decoding) and writing (encoding) are \emph{doubly
dissociable}. \emph{Pure alexia without agraphia} (loss of reading with spared writing, attributable to a disconnection of visual input from the word-processing network) dates to \citet{dejerine1892} and supplies the first half of the
dissociation logic. Its mirror image, \emph{pure agraphia}, isolates a writing deficit
with preserved reading; the writing system has its own architecture, including an
abstract graphemic buffer that holds letter identity and order during spelling
\citep{caramazza1987buffer} and a writing-specific frontal locus (Exner's area)
implicated by direct cortical stimulation \citep{roux2009exner}. Even \emph{decoding}
alone is not monolithic: dual-route models posit parallel lexical and
grapheme--phoneme pathways \citep{coltheart2001drc}. At the same time the two are not
fully independent. Within-subject imaging and lesion evidence indicate a \emph{shared
orthographic core}, with overlapping left ventral occipitotemporal and inferior
frontal representations serving both spelling and reading \citep{purcell2017shared}.
The human picture is two systems with their own failure modes over a partially shared
substrate (precisely the structure we ask whether a decoder-only LLM reproduces or
collapses).

\subsection{Written language as a cultural artifact, and sample efficiency}
\label{sec:related-artifact}

Writing is a recent technology: proto-cuneiform accounting marks appear around
3300~BCE \citep{met_writing}, making literacy a roughly five-millennium-old invention
that each child must be explicitly taught. A striking behavioural consequence is that
phonemic awareness does not arise spontaneously but is a \emph{consequence} of
alphabetic literacy (illiterate adults fail phoneme-manipulation tasks that matched
late-literate adults pass; \citealp{morais1979}). This contrasts sharply with
\emph{spoken} language, acquired universally and without instruction and best
characterized as an evolved instinct \citep{pinker1994}. Our thesis rests on this
asymmetry: an LLM optimizes the cultural artifact, not the evolved faculty.

It does so at a data cost orders of magnitude beyond the human reference. Children are
exposed to on the order of $10^7$--$10^8$ words by adolescence whereas competitive
models consume trillions of tokens; the BabyLM challenge frames this gap and proposes
developmentally plausible budgets \citep{warstadt2023babylm}. \citet{chollet2019measure}
reframes intelligence itself as skill-acquisition efficiency relative to priors and
experience, which makes the human--machine data gap a difference in kind rather than
degree. We treat any human--machine token comparison as illustrative and
order-of-magnitude only: web text is not child-directed speech, and humans receive
multimodal, embodied, interactive input.

\subsection{Transformers, induction heads, and embedding geometry}
\label{sec:related-transformers}

The mechanism LLMs bring to this artifact is general-purpose self-attention
\citep{vaswani2017}, a modality-agnostic routing operation rather than a
language-specific faculty. Mechanistic interpretability has made parts of it legible:
the residual-stream formalism of \citet{elhage2021framework} casts each component as
\emph{reading from} and \emph{writing to} a shared activation stream and identifies how
attention heads compose into \emph{induction heads}, circuits implementing
prefix-matching copy that underwrite much of emergent in-context learning
\citep{olsson2022}. Complementary analyses show heads specialize into interpretable
roles, with a small number doing most of the work \citep{voita2019}.

The geometric heart of our contribution concerns the input embedding $\WE$, which maps
tokens \emph{into} the stream, and the output unembedding $\WU$, which maps the stream
\emph{out} to next-token logits. Our ``reading/writing'' labels track the brain's
input-reading, output-writing convention; note this is the reverse of the
residual-stream usage of \citet{elhage2021framework}, in which components \emph{write
to} and \emph{read from} the stream. We mean input- versus output-side only, and since
$\WE$ and $\WU$ are indexed by BPE tokens rather than graphemes, we reserve
``orthographic'' for the sub-token word-form analysis. \emph{Weight tying} sets
$\WE = \WU^{\top}$, forcing the two codes to be identical
\citep{presswolf2017,inan2017tying}; it is standard in the GPT-2 family and OPT, and is
the architectural condition under which reading and writing are entangled by
construction. The Pythia suite leaves the embeddings \emph{untied}, learning $\WE$ and
$\WU$ separately and exposing aligned checkpoints that let us watch the two codes
differentiate over training \citep{biderman2023}.

The closest prior work to the geometric half of our study is
\citet{lopardo2026weighttying}, which measures alignment between input and output
embeddings and reports that output gradients dominate early training, biasing token
embeddings toward the output space and producing an asymmetric drift. We credit that
work for first reporting this trajectory; our Experiment~7 is a \emph{controlled,
properly nulled replication} on the read/write axis, adding an independent-init floor,
a shuffled-token null, a tied ceiling, and a frequency-decile decomposition that
addresses the gradient-exposure confound. To compare $\WE$ and $\WU$ as
representational \emph{spaces} we use established similarity measures (linear CKA,
\citealp{kornblith2019cka}; orthogonal-Procrustes residual with per-token cosine,
\citealp{smith2017offline}; and mutual $k$-nearest-neighbor overlap,
\citealp{huh2024platonic}), all situated within the representational-similarity
tradition \citep{kriegeskorte2008rsa}. That $\WE$ and $\WU$ should be \emph{coupled but
not identical} in untied models is consistent with evidence that the output embedding
is functionally specialized rather than a transpose of the input embedding, encoding
token-probability and frequency structure \citep{liu2024tokenprob}. Because performance
improves predictably with scale and data \citep{kaplan2020scaling}, we interpret any
geometry along the scale axis against that backdrop and report its trend honestly as
flat and underpowered.

\subsection{Are LLMs a different kind of mind?}
\label{sec:related-minds}

A final strand asks what such systems \emph{are}. Balanced surveys call for an extended
science of intelligence that catalogs distinct modes of understanding rather than
forcing a binary verdict \citep{mitchellkrakauer2023}. A long-running critique holds
that form learned from distribution cannot by itself yield grounded meaning: the
symbol-grounding regress \citep{harnad1990symbol} and the ``stochastic parrot'' framing
\citep{bender2021parrots}. The most directly operationalizable formulation for our
purposes is the dissociation of \emph{formal} from \emph{functional} linguistic
competence: LLMs are strong on formal patterns yet patchy on functional, world-directed
use, paralleling a neural dissociation between the brain's language network and its
reasoning systems \citep{mahowald2024,fedorenko2024language}. We extend this
form/function axis with a second, brain-anchored \emph{read/write} axis. Finally,
psycholinguistics treats comprehension and production as distinct-but-coupled systems
linked by forward models \citep{pickering2013integrated}, and recent work leverages
human production--interpretation asymmetries to probe LLM cognitive plausibility
\citep{lam2025prodinterp}. We attempted the corresponding behavioural test and report
in Section~\ref{sec:res-limits} why it cannot be constructed at the output of a
decoder-only model. The positioning is therefore conservative: we ask not whether LLMs
understand, but whether the way they realize functional literacy (by collapsing the brain's dissociated read/write architecture into one entangled statistical code) places them at a distinct point in the space of possible minds.

\section{Models and Methods}
\label{sec:methods}

Our experiments exploit a single architectural fact. A decoder-only transformer
maps tokens to vectors through an input embedding matrix $\WE \in
\mathbb{R}^{|V| \times d}$ and maps the final residual state back to a
distribution over tokens through an output unembedding (``readout'') matrix
$\WU \in \mathbb{R}^{|V| \times d}$, where $|V|$ is the vocabulary size and $d$
the model width \citep{vaswani2017,elhage2021framework}. We treat $\WE$ as the
model's \emph{reading code} (the input-side token code that turns a token into an
internal vector) and $\WU$ as its \emph{writing code} (the output-side token code
that turns an internal vector into a token). This is an \emph{analogy}, not a
homological claim: $\WE$ and $\WU$ are input- and output-side codings of the same
modality at the same processing stage (indexed by BPE tokens, not graphemes, so
we do not call them ``orthographic'') and not the brain's spatially distinct
reading (ventral occipitotemporal) and writing (fronto-parietal graphemic)
networks. (Our input=reading/output=writing labelling is the reverse of the
residual-stream ``writes-to/reads-from'' convention of
\citet{elhage2021framework}; we intend input- vs output-side only.) The brain's reading and writing
systems doubly dissociate (pure alexia spares writing and pure agraphia spares
reading; \citealp{dejerine1892,caramazza1987buffer,roux2009exner}) while sharing a
partial orthographic core \citep{purcell2017shared}. We ask, by tight analogy
only, whether the LLM's input/output codes are one entangled system or two
dissociable ones, and we never claim the model ``reads'' or ``understands.''

\subsection{Models}
\label{sec:methods-models}

We use two model families that differ in exactly the property of interest:
whether the reading and writing codes are forced to be identical.

\paragraph{Tied (entangled by construction).} The GPT-2 family
(\texttt{distilgpt2}, \texttt{gpt2}, \texttt{gpt2-medium},
\texttt{gpt2-large}; \citealp{radford2019}) and the two OPT models we use
(\texttt{facebook/opt-125m}, \texttt{facebook/opt-350m};
\citealp{zhang2022opt}) tie their embeddings, imposing $\WU = \WE$
\citep{presswolf2017,inan2017tying}. For these models the reading and writing
codes are the same matrix by construction; they serve as a \emph{ceiling}
reference (see \S\ref{sec:methods-eindex}), not as evidence about learning.

\paragraph{Untied (codes learned separately).} The Pythia / GPT-NeoX suite
(\texttt{EleutherAI/pythia-\{14m, 70m, 160m, 410m, 1b, 1.4b\}};
\citealp{biderman2023}) keeps $\WE$ and $\WU$ as independent parameter blocks,
so any geometric similarity between them is learned rather than imposed. Pythia
additionally releases $154$ checkpoints per model, trained on the Pile in a fixed
data order, with $2{,}097{,}152$ tokens per step; this lets us measure how the
codes diverge \emph{over training}, not just at convergence. All models are run
locally on a single RTX 4070.

\subsection{Reading and writing codes, and the entanglement index}
\label{sec:methods-eindex}

To quantify how similar the reading code $\WE$ and the writing code $\WU$ are, we
compare the two $|V| \times d$ matrices with three complementary
representational-similarity measures, each invariant to a different nuisance
transformation, following standard practice in the similarity-of-representations
literature \citep{kriegeskorte2008rsa}.

It helps to picture each matrix as a map that places every vocabulary token at a
point in space, so that $\WE$ and $\WU$ are two maps of the same set of tokens. The
three measures then ask three different questions about whether the maps agree. The
first asks whether they agree on the overall pattern of which tokens sit near which,
ignoring any difference in overall size or orientation. The second asks whether one
map can be rotated bodily onto the other, and how much disagreement is left over
after the best possible rotation. The third ignores global shape entirely and asks a
local question: for each token individually, are its nearest neighbours the same on
both maps? A pair of codes could score high on one and low on another (two maps can
share a global shape while disagreeing token by token), which is why we report all
three rather than any single one.

\begin{enumerate}
  \item \textbf{Linear CKA} (centered kernel alignment) between $\WE$ and $\WU$,
  a scale- and rotation-invariant global similarity of the two row-space
  geometries \citep{kornblith2019cka}; this is our primary metric.
  \item \textbf{Orthogonal Procrustes residual} plus mean per-token cosine: we
  solve for the orthogonal map $R$ minimizing $\lVert \WE R - \WU \rVert_F$ and
  report the residual and the resulting per-token cosine similarity, testing
  whether the two codes are the same space up to a rotation/reflection
  \citep{smith2017offline}.
  \item \textbf{Mutual $k$-nearest-neighbor overlap}, the fraction of each
  token's $k$ nearest neighbors in $\WE$ that are also neighbors in $\WU$, a
  local/neighborhood-structure measure \citep{huh2024platonic}.
\end{enumerate}

We fuse these into a single composite entanglement index $\Eindex \in [0,1]$.
Writing $c$ for linear CKA, $r$ for the Procrustes residual, $\bar\theta$ for the
mean per-token cosine, and $o$ for mean mutual-$k$NN overlap ($k=10$), and using
the tied-model ceiling and the independent-init floor (below) to normalize each
component to $[0,1]$ via $\tilde m = \mathrm{clip}\!\big((m-m_{\text{floor}})/
(m_{\text{ceil}}-m_{\text{floor}}),0,1\big)$ (with the residual oriented as
$1-r$), we define $\Eindex$ as the unweighted mean of the four normalized
components,
\[
\Eindex \;=\; \tfrac14\big(\tilde c + \widetilde{(1-r)} + \tilde{\bar\theta} + \tilde o\big),
\]
so $\Eindex=1$ at the tied ceiling and $\Eindex=0$ at the independent-init floor.
The two anchors are what make the number mean anything, in the way a temperature
scale needs two fixed points before a reading is informative: a raw similarity value
cannot be called high or low until one knows what fully unrelated codes score (the
floor) and what deliberately identical codes score (the ceiling).
We always report the raw components alongside $\Eindex$ so the composite never
hides a divergent component, and we verified the trends are not an artifact of the
equal-weighting choice. To make $\Eindex$ interpretable we anchor it against three
controls:

\begin{itemize}
  \item a \textbf{tied ceiling}: for tied models $\WU = \WE$, so $\Eindex = 1$
  exactly (a sanity check on the pipeline, \emph{not} an empirical finding);
  \item an \textbf{independent-initialization floor}: $\Eindex$ between two
  freshly, independently initialized matrices, the value expected with no shared
  learning (empirically near zero on every component);
  \item a \textbf{shuffled-token null}: $\Eindex$ recomputed after randomly
  permuting the token-to-row correspondence between $\WE$ and $\WU$, which
  destroys any token-aligned structure while preserving each matrix's marginal
  geometry.
\end{itemize}

For the training-trajectory analysis (\S\ref{sec:methods-probes}, Exp.~7) we
additionally decompose $\Eindex$ and the per-matrix drift by token
\textbf{frequency decile}, to separate genuine read/write differentiation from
the confound that frequent tokens simply receive more gradient updates on the
output side. All headline similarity values are reported with bootstrap $95\%$
confidence intervals (2{,}000 resamples, percentile method); the
independent-init floor is computed per model from two Gaussian matrices matched
to each code's scale under a fixed seed, and mutual-$k$NN uses $k=10$ with query
subsampling for tractability on the full vocabulary. We note one interpretive
caveat inherited from this design: $\WE$ and $\WU$ are compared in a common token
basis, so the metrics are sensitive to the byte-pair tokenization shared by the
two codes rather than to any sub-symbolic orthography.

\subsection{The ten probes}
\label{sec:methods-probes}

The ten experiments trace a single arc: the first six (Exps.~1--6) consolidate
established results, Exp.~7 introduces the read/write analysis in the weights, and
Exps.~8--10 are the three attempts to test it at the output. Self-attention is the general
mechanism (Exp.~1) that acquires written language as a cultural overlay (Exp.~2)
at a large data cost (Exp.~3), mastering form before meaning (Exp.~4) by fitting
and storing text statistics (Exps.~5--6); on top of this we ask whether the model
carries a single entangled read/write code (Exp.~7) and whether it behaves coupled
rather than doubly dissociated (Exp.~8).

\paragraph{Exp.~1: Induction heads and causal ablation.} We localize
prefix-matching induction heads \citep{olsson2022,elhage2021framework} in all
models via an induction score, then causally ablate the top-scoring head and
measure the change in in-context learning (ICL) on a synthetic copy-completion
task, benchmarked against a null distribution of $24$ random-head ablations
(so the smallest attainable permutation $p$ is $1/25\approx0.04$) and
against MLP-block ablations \citep{voita2019}. The claim is
scoped to ``induction circuits are causally implicated in ICL,'' not to
attention dominating MLPs.

\paragraph{Exp.~2: Orthographic word-form acquisition.} We track the
orthographic word-form gap (bits-per-byte on illegal minus real letter strings)
across Pythia checkpoints, test its temporal alignment with the induction phase
change, run a causal test (does ablating the top induction head remove the gap?),
and fit a layerwise linear probe for the real/legal/illegal three-way
distinction against a permutation null, with BPE-confound controls (token-count
and subword-unigram matching). Because a from-scratch transformer has no
pre-existing evolved substrate, the analogy to neuronal recycling and the VWFA
\citep{dehaene2007recycling,dehaene2011vwfa,morais1979} is to acquisition
\emph{timing} only.

\paragraph{Exp.~3: Tokens-to-competence.} Using the released Pythia
checkpoints (tokens $=$ step $\times 2{,}097{,}152$), we plot loss and BLiMP
grammaticality (a benchmark of minimal sentence pairs differing in one grammatical
feature, scored by whether the model assigns the grammatical member higher
probability) against tokens seen and compare the data budget to the
$\sim$100M-word ``by age 13'' human reference
\citep{warstadt2023babylm, kaplan2020scaling}. This comparison is
illustrative and order-of-magnitude only: humans receive multimodal, embodied,
interactive input, and web text is not child-directed speech, so this is not a
claim of cognitive equivalence \citep{chollet2019measure}.

\paragraph{Exp.~4: Formal vs.\ functional competence.} We contrast formal
linguistic competence (BLiMP, with a per-phenomenon breakdown) against functional
competence (PIQA, a templated situation-tracking probe, and absurdity judgments),
with difficulty/length-matched controls and a surface baseline built from positive
pointwise mutual information (PPMI, a co-occurrence statistic that captures how
often two words appear together relative to chance, and therefore what can be
predicted from word co-occurrence alone) under two normalization schemes, operationalizing the form/function dissociation of
\citet{mahowald2024,fedorenko2024language}.

\paragraph{Exp.~5: Compression and the statistical artifact.} We fit the
bits-per-byte ($\bpb$) scaling law in non-embedding parameters, build a BPE
Kneser--Ney $n$-gram comparator over the \emph{same} tokenization, measure
surface-perturbation fragility,
and test whether compression decouples from functional competence across the six
Pythia sizes. The decoupling sub-claim is reported as a planned test that may
fail.

\paragraph{Exp.~6: Repository vs.\ grounding ceiling.} We measure extractable
verbatim memorization vs.\ scale \citep{carlini2023quantifying}, an
imitation-vs-innovation gap on three-choice tool items against a rarity baseline
\citep{yiu2024imitation}, and a bounded symbol-grounding demonstration comparing
hidden-state geometry to perceptual norms against a PPMI baseline
\citep{harnad1990symbol}. The three parts measure different constructs;
``repository, not abstraction'' is a hypothesis, not proven by any single probe.

\paragraph{Exp.~7: The entanglement index across architecture, scale, and
training.} Using the $\Eindex$ machinery of \S\ref{sec:methods-eindex}, we
compare $\WE$ and $\WU$ (i) across tied vs.\ untied architectures, (ii) across the
Pythia scale ladder, and (iii) along the Pythia-160m (and 410m, 14m) training
trajectory. The trajectory analysis is a controlled, properly-nulled replication
of the asymmetric output-side drift first reported by
\citet{lopardo2026weighttying}, using the independent-init floor, shuffled-token
null, and frequency-decile controls defined above
\citep{presswolf2017, inan2017tying, smith2017offline}.
Summary values appear in Table~\ref{tab:exp7_entanglement}.

\paragraph{Exps.~8--10: the behavioural and causal tests, and why they are
withdrawn.} We attempted three successive operationalizations of the read/write
distinction at the output, and report their designs here because the reasons they
fail are results in their own right (Section~\ref{sec:res-limits}).

\emph{Exp.~8, a contingency table.} Comprehension is log-probability
discrimination (does the model prefer the correct written form over a competitor?)
and production is greedy generation (does the model write the form?). These are
mechanistically distinct measurements, so we assembled the $2\times2$ table whose
off-diagonal ``dissociation mass'' would be the analogue of the brain's
alexia/agraphia double dissociation \citep{coltheart2001drc,
pickering2013integrated,lam2025prodinterp,mahowald2024}, summarized the coupling by
$\phi$ (the Matthews correlation between comprehension- and production-correctness),
and tested a geometry$\rightarrow$behaviour bridge by correlating $\Eindex$ with
$\phi$. Production's success set is a subset of comprehension's, however, and for
single-token pairs that nesting is a theorem; in our stimulus set the two marginals
were additionally measured on disjoint item subsets, making the off-diagonal an exact
function of the margins. The table, $\phi$, the sign test over models, and the bridge
computed against $\phi$ are all withdrawn.

\emph{Exp.~9, an architecture contrast.} To ask whether coupling is specific to the
decoder-only forward path we ran a matched probe across decoder-only (Pythia),
encoder--decoder (T5~v1.1 \citep{raffel2020t5}, plus the Pile-trained Pile-T5 as a
data-matched control) and masked-encoder (BERT \citep{devlin2019bert}; RoBERTa
\citep{liu2019roberta}) models, and measured the representational separation of the
encoder and decoder stacks as $1-\mathrm{CKA}$ between their per-layer hidden states
on the same inputs. The behavioural half inherits Exp.~8's defect. The
representational half we withdraw for three independent reasons: the contrast is
fixed by architecture class before any data is collected, the measure is unchanged by
destroying linguistic structure, and its value is unstable in the stimulus set. We
re-measured it on hundreds of distinct corpus sentences with a random-init floor, a
shuffled-token null, and a within-stack depth-drift reference; details and numbers
are in Section~\ref{sec:res-limits}.

\emph{Exp.~10, a lesion contrast.} Because baseline nesting constrains the two
success sets but not their derivatives, we ablated one attention head, attention
sublayer or MLP sublayer at a time and asked whether comprehension and production are
damaged differentially, scoring both on graded capabilities (a comprehension margin,
and a production capability that is zero exactly when greedy decoding would write the
form) rather than thresholded accuracies. Severity is controlled with the mean
Kullback--Leibler divergence of the ablated next-token distribution from its
unablated value at the target's positions, which is neutral between reading and
writing. Inference is a paired bootstrap over items with a step-down max-statistic
permutation test over sites. The per-site index is not scale-identified and the
pooled coupling statistic is explained by a term the two scores share; both are
withdrawn, with the diagnosis reported in Section~\ref{sec:res-limits}. Throughout,
we read these attempts through the ``space of possible minds'' lens
\citep{sloman1984,shanahan2025,mitchellkrakauer2023}.

\subsection{Reproducibility and honesty}
\label{sec:methods-repro}

All experiments run in fp32 by default: we observed that fp16 produces NaNs on
the GPT-NeoX/Pythia models, and bf16 requires \texttt{torch} $\geq 2.1$. We use
fixed random seeds and frozen stimulus sets (including the author-made
situation-tracking and tool-innovation items) so that every reported number is
reproducible on a single RTX 4070. All models and benchmarks we use are publicly
available (the GPT-2, OPT, and Pythia/GPT-NeoX checkpoints, and the Pile, BLiMP,
and PIQA), so every result can be regenerated from public sources following the
procedures described above.

\paragraph{Statistical approach.} Our inferences are deliberately modest. With
six model sizes and three model families (sharing tokenizers within family), the
cross-model and scale analyses rest on small, non-independent samples; we treat
all fits as \emph{descriptive}, apply no multiple-comparison correction, and
designate no test as confirmatory at a corrected $\alpha$. Where a sign or
permutation test is reported (e.g.\ the behavioral coupling), we state it
explicitly; elsewhere we attach bootstrap $95\%$ confidence intervals and label
small-$n$ relationships ``suggestive.'' Model family is a confound for every
cross-model comparison, which we flag rather than correct away. Finally, because
a decoder-only architecture commits to a single forward path, some read/write
coupling is expected a priori (our claims concern its magnitude and trajectory,
not its existence).

\section{Results}
\label{sec:results}

We organize the results along the argument arc. Self-attention is a
modality-general routing primitive, and induction circuits within it are
causally implicated in in-context learning (Section~\ref{sec:res-mechanism}); on
written language the model acquires an orthographic word-form code as a cultural
overlay (Section~\ref{sec:res-acquisition}) at a data cost orders of magnitude
beyond the human budget (Section~\ref{sec:res-efficiency}); it masters form
before meaning (Section~\ref{sec:res-form}) by compressing and storing text
statistics (Sections~\ref{sec:res-compression}--\ref{sec:res-repository}); and it
does so with a single, entangled read/write code
(Section~\ref{sec:res-entanglement}). Because a decoder-only model has a single
forward path, \emph{some} read/write coupling is expected a priori; our
contribution is therefore not the existence of coupling but its
\emph{quantification} (its distance from an independent-init floor and a tied
ceiling, and the shape of its training trajectory).

We then report a second, negative contribution. We attempted the matching
behavioural test, that comprehension and production should fail together rather
than dissociate, and neither it nor two successive redesigns could be made valid
(Section~\ref{sec:res-limits}). The obstacles are structural rather than
incidental: each candidate measurement reduced to an algebraic identity between
the two scores or to a quantity with no identified scale. We state the three
results explicitly, because in a model whose reading and writing are read off a
single next-token distribution, the fact that no pair of output-side scores
isolates one ability from the other is itself informative about entanglement. All
confidence intervals are bootstrap 95\% intervals unless noted; the
compression/reasoning decoupling and several secondary trends are reported as
plain nulls.

\subsection{Induction circuits are causally implicated in in-context learning}
\label{sec:res-mechanism}

In-context copying is supported by induction heads in every model we examine.
Prefix-matching induction scores are high across all eight small open models,
ranging from $0.869$ (pythia-70m) to $0.978$ (pythia-160m), far above the
$\sim$$0.011$ chance rate expected for a random head
(Table~\ref{tab:exp1_induction}; Appendix Figure~\ref{fig:exp1_heatmap}). To test
whether these heads are \emph{causally} implicated in in-context learning (ICL)
rather than merely correlated with it, we zero-ablate the single top induction
head and measure the change in a synthetic copy-completion ICL score relative
to a 24-random-head null. The effect is significant for opt-125m ($z=9.57$),
pythia-160m ($z=4.25$), and pythia-410m ($z=2.07$) (all permutation
$p=0.04$, which is the resolution floor $1/25$ of the 24-head null, not a precise
estimate); it is weak or non-significant in the GPT-2 family. Crucially, we
also ablate the MLP block at the same layer: its effect on ICL is comparable to
or larger than the head's (e.g.\ pythia-70m $\Delta\mathrm{ICL}=2.16$ for the
MLP vs $1.41$ for the head). We therefore scope the claim narrowly:
\emph{induction circuits are causally implicated in ICL}, not that attention
dominates the MLP. Across models the relationship between maximum induction
score and in-context gain is positive but does not reach significance
(Spearman $\rho=0.57$, $p=0.139$, $n=8$; Appendix Figure~\ref{fig:exp1_scatter}). We
report this small-$n$ cross-model trend as suggestive.

This mechanism is not specific to language: self-attention is a general
information-routing operation \citep{vaswani2017,elhage2021framework} rather
than a language-specific faculty. What follows is what that general
mechanism does when its training signal is \emph{written} language.

\subsection{Written language as an acquired cultural overlay}
\label{sec:res-acquisition}

\begin{figure}[t]
  \centering
  \includegraphics[width=\linewidth]{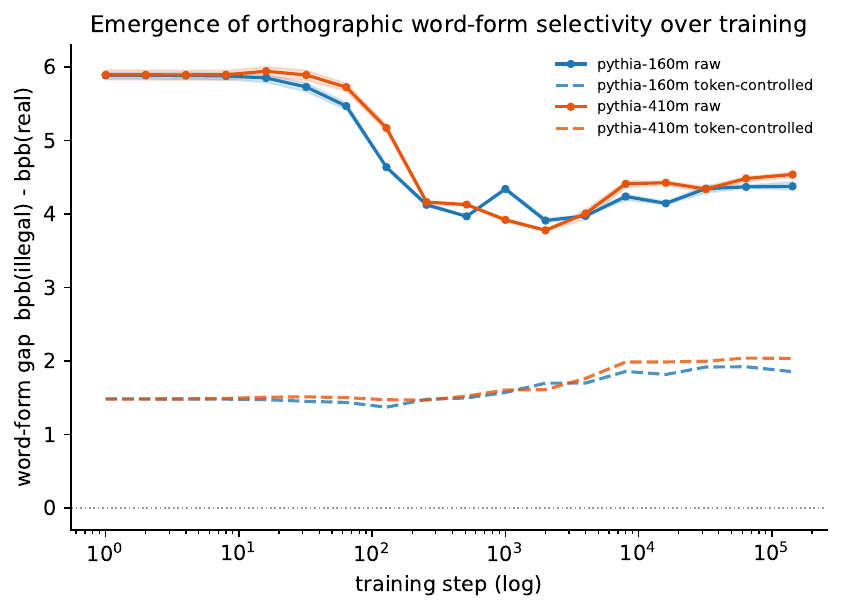}
  \caption{Orthographic word-form gap ($\bpb(\text{illegal})-\bpb(\text{real})$)
  over Pythia training. The raw gap (dominated by a BPE tokenization confound) is
  large at initialization and shrinks; the token-count--controlled gap, the
  genuine acquired signal, grows with tokens seen.}
  \label{fig:exp2_gap}
\end{figure}

A from-scratch transformer has no evolved substrate for orthography; any
sensitivity to legal letter sequences must be \emph{acquired}. Using the aligned
Pythia checkpoints, we measure an orthographic word-form gap, defined as
$\bpb(\text{illegal})-\bpb(\text{real})$ over the continuation tokens of a word
form (carrier masked), ordered real $<$ legal-pseudoword $<$ illegal at every
checkpoint. The \emph{raw} gap is dominated by a tokenization confound (illegal
consonant strings fragment into many more BPE sub-tokens than real words), so for
pythia-160m it is already large at random initialization ($5.88$ bits/byte) and,
if anything, \emph{shrinks} as the model learns to compress all three classes
(to $4.38$ at the final checkpoint). The genuine, acquired signal appears only
once this confound is removed: the token-count--residualized gap \emph{grows}
over training from $1.48$ to $1.85$ bits/byte, and a token-count--matched subset
grows from $2.93$ to $3.80$ (Table~\ref{tab:exp2_wordform_dynamics};
Figure~\ref{fig:exp2_gap}). Orthographic word-form selectivity is therefore
acquired over training, as the neuronal-recycling framing predicts for a cultural
skill, though we cannot fully exclude residual subword-frequency structure. We do
\emph{not} claim a clean co-onset with the induction phase change: the raw-gap
onset is itself a tokenization artifact (its maximum-slope step is step~1), and,
as we show next, the induction circuit is causally unrelated to the gap. We thus
report the controlled word-form gap as a \emph{gradually acquired} code rather
than one temporally locked to circuit emergence.

\begin{figure}[t]
  \centering
  \includegraphics[width=\linewidth]{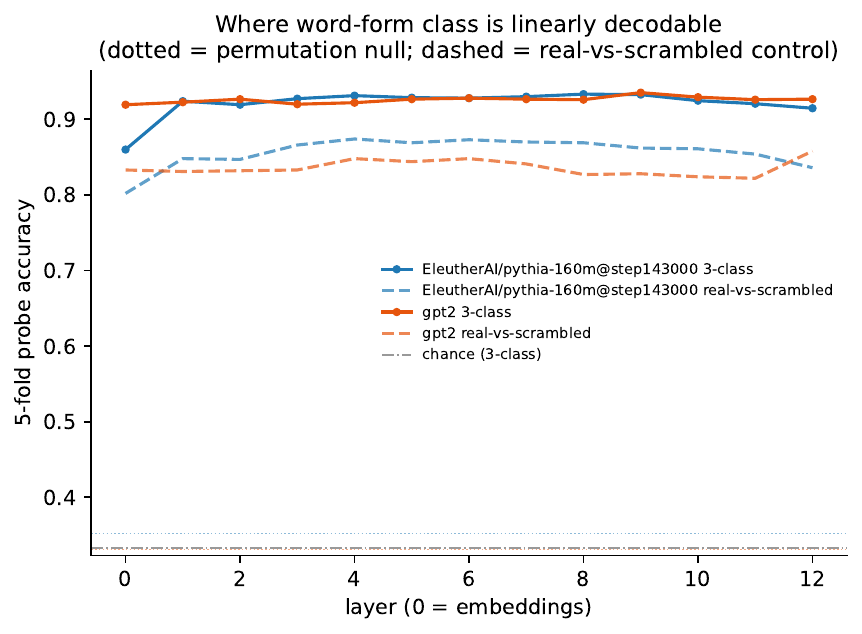}
  \caption{Layerwise linear-probe decodability of word-form class. Decodability
  peaks in mid-to-late layers above the permutation null.}
  \label{fig:exp2_probe}
\end{figure}

We confirm the same orthographic distinction directly. A linear probe
separates real/legal/illegal word forms with $0.933$ three-class accuracy at
pythia-160m's best layer (layer 8), well above both chance ($0.333$) and a
permutation null ($0.352$), and a token-count--matched real-vs-scrambled
control ($0.869$) rules out token-count leakage
(Table~\ref{tab:exp2_probe}). Decodability peaks in mid-to-late layers, the
``letterbox'' depth profile (Figure~\ref{fig:exp2_probe}).

Critically, this orthographic code is \emph{not} carried by the induction
circuit. Ablating the top induction head (L4H6 in pythia-160m) leaves the
word-form gap essentially intact: it moves from a clean $4.324$ to $4.361$
under ablation, a fraction remaining of $1.009$ (a matched random-head control
shifts the gap by only $-0.001$; the gap does not collapse). This is an honest
negative: the mechanism that supports in-context copying (Section~\ref{sec:res-mechanism})
is dissociable from the one that encodes orthographic legality. We emphasize
that the analogy to the visual word form area and neuronal recycling
\citep{dehaene2007recycling,dehaene2011vwfa,morais1979} is to \emph{acquisition
timing} only; a randomly initialized transformer has no pre-existing evolved
cortex to recycle.

\subsection{The data cost of competence}
\label{sec:res-efficiency}

Both loss and grammaticality improve smoothly as a function of tokens seen
(Table~\ref{tab:exp3_data_dynamics}). For
pythia-160m, perplexity on held-out text falls from $\sim$$6.3\times10^{4}$ at
step~1 to $\sim$$50$ at the final checkpoint, while BLiMP grammaticality rises
from chance ($0.484$) to $0.744$; the larger pythia-410m reaches $0.805$. An
across-size loss-vs-$N$ power law at the final checkpoint has exponent
$\alpha_N=0.100$ (with the within-run fits flagged as autocorrelated, so their
intervals understate uncertainty). The machine reaches this competence only
after $\sim$$3\times10^{11}$ tokens ($\sim$$2.3\times10^{11}$ word-equivalents
assuming a nominal $1.3$ tokens/word), three to four orders of magnitude beyond the
$\sim$100M-word budget a child has by age 13 (band 50--200M words, i.e.\
65--260M tokens; \citealp{warstadt2023babylm}). We treat the human comparison as
illustrative and order-of-magnitude only: children receive multimodal,
embodied, interactive, child-directed input, whereas the Pile is adult written
text, so this is not a confirmed cognitive equivalence
\citep{chollet2019measure}.

\subsection{Form before meaning}
\label{sec:res-form}

Across the Pythia scale ladder, formal competence (BLiMP grammaticality)
exceeds and saturates ahead of functional competence
(Table~\ref{tab:exp4_form_function}). At the top of
our ladder, pythia-1.4b reaches BLiMP $0.824$ [$0.807$, $0.841$] while its
functional composite (the mean of the PIQA physical-commonsense,
situation-tracking, and absurdity-detection accuracies) reaches
$0.757$, a form$-$function gap of $0.067$. Form rises early and plateaus;
function lags and remains below it. These results hold under a
difficulty/length-matched control and a PPMI surface baseline, following the
formal/functional distinction of \citet{mahowald2024,fedorenko2024language}. We
are candid that the gap is modest at 1.4b, that the situation-tracking items are
author-made and templated (frozen across runs), and that there is no human
ceiling; the claim is scoped to these probes, not to meaning in general
\citep{harnad1990symbol}.

\subsection{Optimizing a statistical artifact: compression and its limits}
\label{sec:res-compression}

Compression of held-out text improves as a clean power law in non-embedding
parameters: bits/byte falls from $1.49$ (pythia-14m) to $0.82$ (pythia-1.4b)
with exponent $\alpha=0.090$ [$0.080$, $0.107$] and $R^2=0.984$ (no-offset fit;
Table~\ref{tab:exp5_scaling}), consistent with
established scaling laws \citep{kaplan2020scaling}.

Compared against a Kneser-Ney BPE $n$-gram on the \emph{same} tokenization, the
interpretable signal is the scaling of the neural$-$$n$-gram gap, not the raw
overlap. The neural/$n$-gram perplexity ratio falls from $0.221$ (pythia-14m)
to $0.025$ (pythia-1.4b): the neural model increasingly outpaces the $n$-gram
with scale (Table~\ref{tab:exp5_ngram}). Top-1 next-token agreement between the
$n$-gram and pythia-160m is only $0.290$, with a symmetric KL of $3.03$ nats, so
we frame the result around the \emph{scaling of the gap} rather than claiming
the $n$-gram recovers most of the model.

The neural model is also consistently more fragile to word-order corruption
than the $n$-gram: the model$-$$n$-gram excess $\Delta$NLL under a length-5
window shuffle grows with scale, from $0.96$ nats (pythia-14m) to $1.66$ nats
(pythia-1.4b) (Table~\ref{tab:exp5_perturbation}). Zipf reproduction is demoted
to a descriptive observation, since its own shuffled control shows it to be
near-trivial.

\begin{figure}[t]
  \centering
  \includegraphics[width=\linewidth]{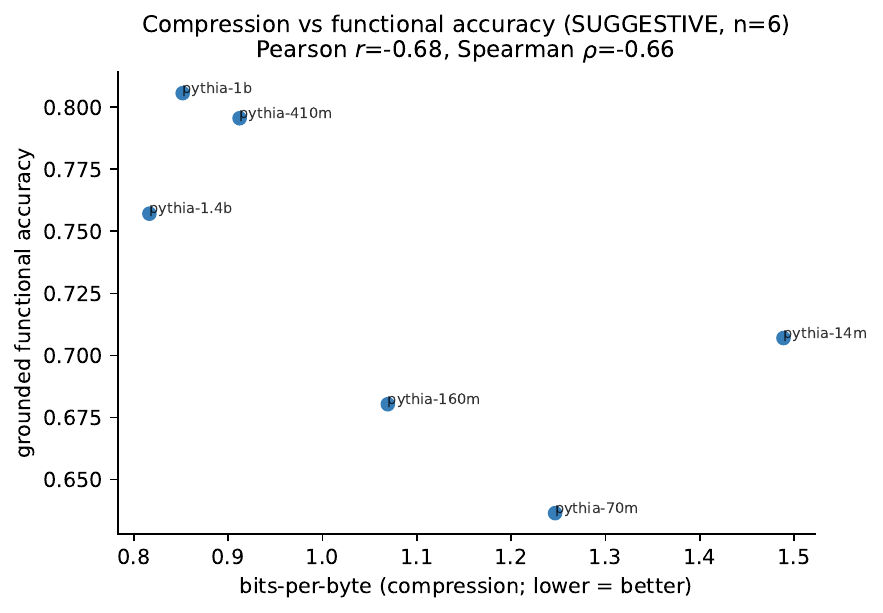}
  \caption{Compression (bits/byte) vs functional competence across six sizes.
  The point estimate runs opposite to decoupling (Pearson $r=-0.68$, $p=0.14$,
  $n=6$); underpowered, but the decoupling hypothesis is not supported.}
  \label{fig:exp5_decoupling}
\end{figure}

\paragraph{Null: no evidence that compression decouples from reasoning.} We
tested whether better compression and better functional competence
\emph{dissociate} across scale. We find no evidence for decoupling: across the
six Pythia sizes the point estimate runs in the \emph{opposite} direction
(bits/byte vs functional composite Pearson $r=-0.68$, $p=0.14$; Spearman
$\rho=-0.66$, $p=0.16$; $n=6$; Figure~\ref{fig:exp5_decoupling}), i.e.\ if
anything the two improve together. With $n=6$ this is underpowered and cannot
establish co-variation either; the only safe conclusion is that the
``compression decoupled from reasoning'' hypothesis is \emph{not supported}, and
we withdraw it as a positive claim.

\subsection{A knowledge repository against a grounding ceiling}
\label{sec:res-repository}

\begin{figure}[t]
  \centering
  \includegraphics[width=\linewidth]{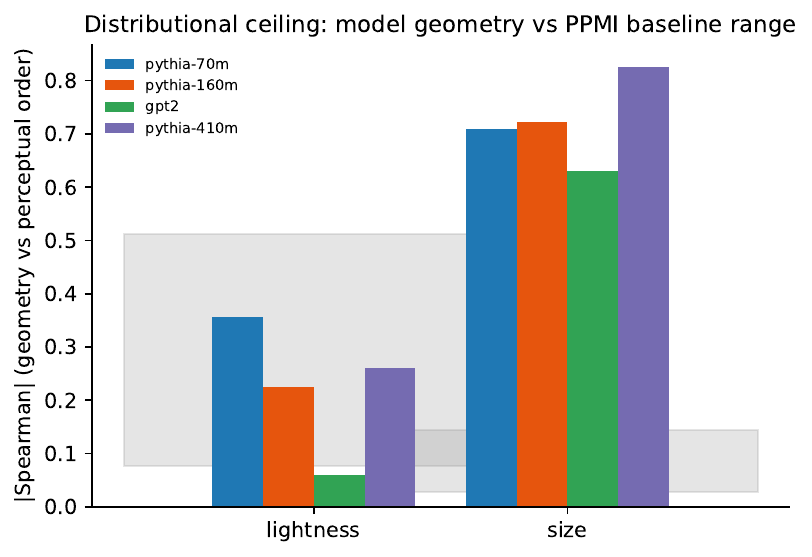}
  \caption{Hidden-state geometry vs perceptual norms against a PPMI baseline.
  Size is captured above the baseline; lightness is not.}
  \label{fig:exp6_grounding}
\end{figure}

Three separate probes characterize the model as a text repository operating
against a distributional ceiling; we report them separately rather than merged
(Table~\ref{tab:exp6_repository}). First, extractable verbatim memorization of
Pile text grows log-linearly with scale, with slope $0.011$ [$0.007$, $0.017$]
in $\log$ non-embedding parameters ($R^2\approx0.63$), against a non-member
negative control consistent with prior
memorization laws \citep{carlini2023quantifying}. Second,
on three-choice tool-use items, the imitation$-$innovation gap reaches $0.090$
while a PPMI rarity baseline reproduces a gap of $0.000$, so the innovation
deficit is not explained by option rarity \citep{yiu2024imitation}. Third, on a
bounded symbol-grounding demonstration, hidden-state geometry tracks perceptual
\emph{size} norms with $|\rho|$ up to $0.83$ (rising with scale), beating a PPMI
baseline ($|\rho|$ max $0.51$), whereas for \emph{lightness} the model does
\emph{not} beat the PPMI baseline (Figure~\ref{fig:exp6_grounding};
\citealp{harnad1990symbol}). The three probes measure
different constructs, and ``repository, not abstraction'' is a hypothesis, not
proven by any single one.

\subsection{A single, entangled read/write code}
\label{sec:res-entanglement}

We now turn to the central claim. We treat the input embedding $\WE$ as a
``reading code'' (the input-side token code) and the output unembedding
$\WU$ as a ``writing code'' (the output-side token code); we use ``reading/
writing'' for the input/output side only, and reserve ``orthographic'' for the
sub-token word-form analysis of Section~\ref{sec:res-acquisition}, since $\WE$
and $\WU$ are indexed by BPE tokens, not graphemes. We quantify their
entanglement with a composite index $\Eindex\in[0,1]$ fusing linear CKA
\citep{kornblith2019cka}, an orthogonal-Procrustes residual with per-token
cosine \citep{smith2017offline}, and mutual $k$-NN overlap
\citep{huh2024platonic}, against an independent-init floor, a shuffled-token
null, and a tied ceiling \citep{kriegeskorte2008rsa}.

\begin{figure}[t]
  \centering
  \includegraphics[width=\linewidth]{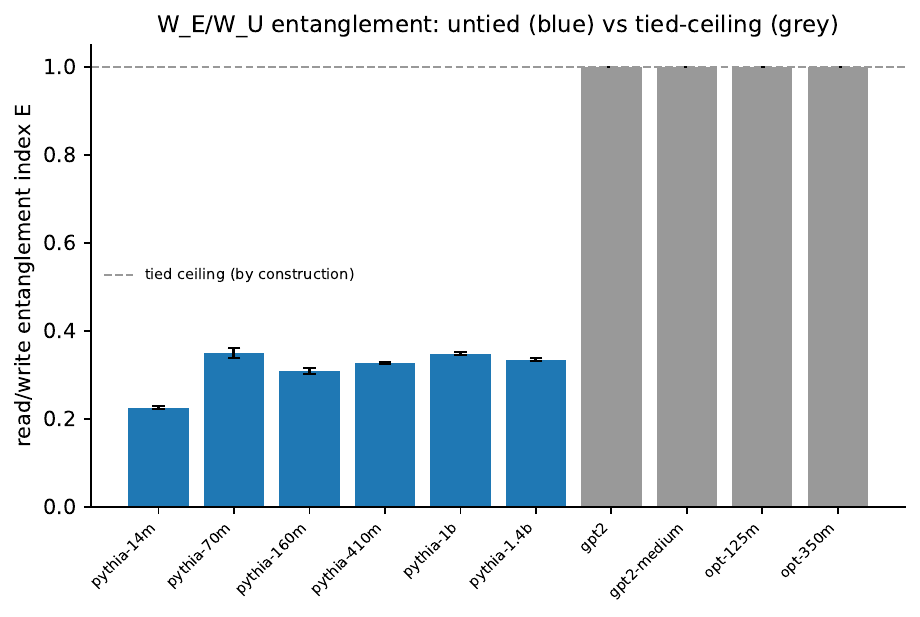}
  \caption{Entanglement index $\Eindex$ by architecture. Tied models sit at the
  ceiling by construction; untied Pythia models sit well below it and far above
  the independent-init floor.}
  \label{fig:exp7_arch}
\end{figure}

The tied models (gpt2, gpt2-medium, opt-125m, opt-350m) return $\Eindex=1.000$
\emph{by construction} ($\WE=\WU^{\top}$;
\citealp{presswolf2017,inan2017tying}); this is a pipeline sanity check, not a
finding. The untied Pythia models are the substance: $\Eindex$ ranges from
$0.226$ to $0.349$, with every component measure far above its independent-init
floor (Table~\ref{tab:exp7_entanglement}; Figure~\ref{fig:exp7_arch}). In plain terms, the two codes sit roughly a third of the way from ``completely
unrelated'' to ``forced to be identical''. The reading and
writing codes are thus \emph{coupled but sub-ceiling}: one entangled code, not
the two dissociated systems of the brain, and also not a trivial collapse onto a
single shared matrix. The coupling is genuine per-token structure rather than a
global-subspace artifact: the observed per-token cosine and mutual-kNN overlap
greatly exceed both a shuffled-token null and the independent-init floor in every
untied model (Table~\ref{tab:exp7_nulls}). The sub-ceiling differentiation is
moreover consistent with evidence that the output embedding is functionally
specialized rather than a mirror of the input embedding
\citep{liu2024tokenprob}.

\begin{figure}[t]
  \centering
  \includegraphics[width=0.8\linewidth]{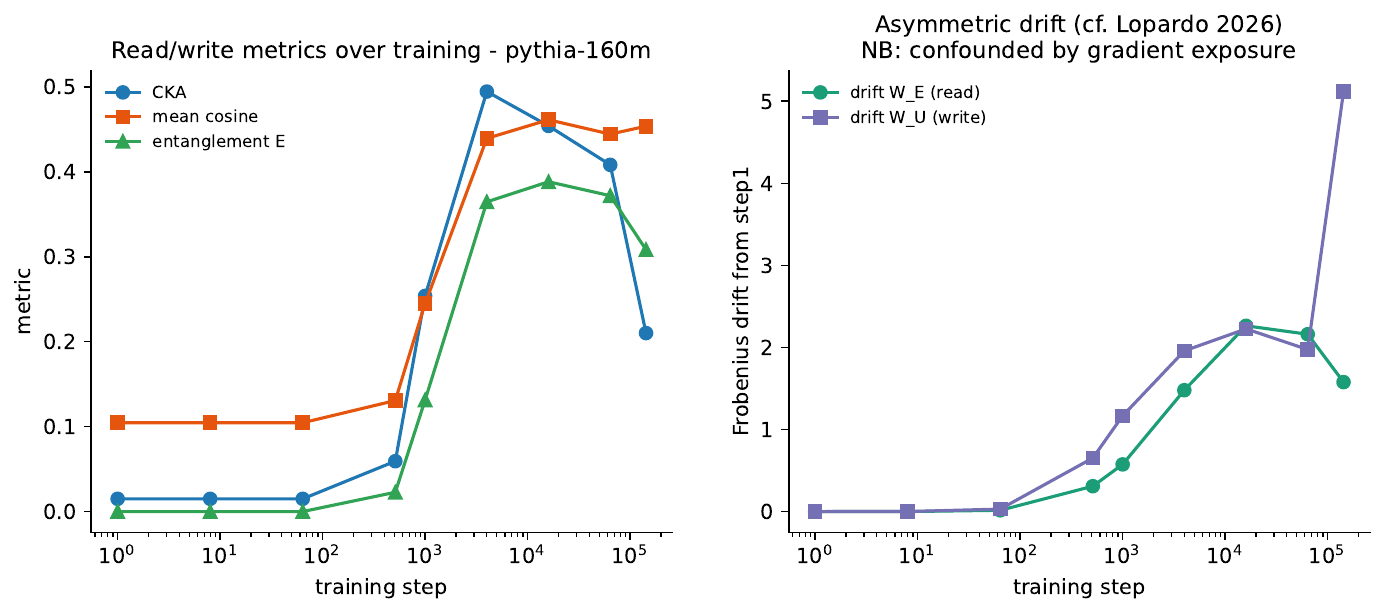}
  \caption{Couple-then-differentiate trajectory of $\WE$/$\WU$ CKA over
  pythia-160m training: rise to a peak of $0.494$ at step~4000, then partial
  differentiation to $0.210$ by step~143000, with asymmetric writing-code drift.
  The asymmetric drift replicates \citet{lopardo2026weighttying}.}
  \label{fig:exp7_dev}
\end{figure}

The training trajectory is non-monotonic and informative
(Figure~\ref{fig:exp7_dev}). For pythia-160m, $\WE$/$\WU$ CKA starts at $0.015$
(initialization, effectively independent), rises to $0.254$ by step~1000, peaks
at $0.494$ at step~4000, then \emph{differentiates} back down to $0.408$ at
step~64000 and $0.210$ at the final step~143000: the codes couple early and
then partly pull apart. This differentiation is asymmetric: at step~143000
the writing code has drifted $\sim$$3.2\times$ as far as the reading code from
initialization (drift $\WU=5.12$ vs drift $\WE=1.58$). This asymmetric-drift
trajectory was \emph{first reported by} \citet{lopardo2026weighttying}; our
contribution here is a controlled replication on the read/write axis, with an
independent-init floor, a shuffled-token null, and a frequency-decile
decomposition addressing the gradient-exposure confound. That confound (the
worry that $\WU$ simply ``moves more'' because the output matrix receives a
softmax gradient on every step while input rows update only for tokens seen) is
ruled out by decomposing the drift by token-frequency decile: the writing code
drifts roughly $3\times$ as far as the reading code in \emph{every} decile, from
the rarest to the most frequent tokens (Table~\ref{tab:exp7_decile_drift}), so
the asymmetry is not carried by frequent tokens alone.

\paragraph{Null: no entanglement trend with scale.} Across the six untied
sizes, $\Eindex$ does not vary systematically with model size: Spearman$(\Eindex,
\log N)=0.37$ (permutation $p=0.49$, $n=6$; Appendix Figure~\ref{fig:exp7_scale}). We
report this as a flat, underpowered null rather than a trend.

\subsection{The read/write distinction resists operationalization at the output}
\label{sec:res-limits}

Having found one coupled read/write code in the weights, we set out to test the
matching behavioural prediction: that comprehension and production should fail
together rather than dissociate. We report here that this test, and two successive
redesigns of it, could not be made valid. The obstacles are not implementation
defects but properties of what the measurements are: each attempt reduced to an
algebraic identity between the two scores, or to a quantity with no identified
scale. We give the three results because they are, we think, the substantive
finding of this half of the paper. In a model whose reading and writing are read
off one next-token distribution, no pair of output-side scores isolates one ability
from the other, and \emph{that} is a form of entanglement no index had to be
constructed to see.

\paragraph{Result 1: for minimal pairs the LLM-alexia cell cannot fire.}
The natural analogue of the alexia/agraphia contrast is a
comprehension~$\times$~production contingency table, whose
``produces-but-does-not-comprehend'' cell would be the model's pure alexia.
That cell is empty by logic. Production, operationalized as greedy generation of the
target form~$F$, is a $|V|$-way task; comprehension, operationalized as ranking $F$
above one competitor $F'$, is $2$-way. If greedy decoding emits $F$, then $F$'s
token was the argmax over the entire vocabulary, so $F$ necessarily outranks $F'$:
production's success set is a subset of comprehension's. For a one-token-versus-
one-token pair this is a theorem, not a tendency. The cell can fire only on pairs
that have equal token counts \emph{and} at least two tokens, but equal token
count is exactly what makes a per-token mean log-probability comparison unbiased,
so the two requirements compete. On a purpose-built $50$-item pool, one item
satisfied both under the GPT-2 and Pythia tokenizers simultaneously; most pairs
were single-token, the remainder length-mismatched. Adding tokenizers makes it
worse, and the only remaining route to a non-empty cell is to let $F'$ beat $F$ on
the \emph{mean} through length normalization, which would manufacture the result
from a scoring artifact.

In our original stimulus set the same degeneracy took a second and more concrete
form. Comprehension saturated at $1.000$ on one item family while production
floored at $0.000$ on the other, so the two marginals were measured on
\emph{disjoint} item subsets and the off-diagonal mass reduced exactly to the
difference of the two marginals. We verified that identity to numerical precision
on every model~$\times$~checkpoint row we had computed. A contingency table whose
cells are exact functions of its margins cannot test independence, so the
coupling statistic, its sign test, and the geometry$\rightarrow$behaviour bridge
built on top of it are all withdrawn.

\paragraph{Result 2: differential-damage indices are not scale-identified.}
Baseline nesting constrains the two success \emph{sets} but not their
\emph{derivatives}, so we turned to the model analogue of a lesion study: ablate
one component at a time and ask whether comprehension and production are damaged
differentially. This requires comparing two damages on a common scale, and no such
scale exists here. The damage distributions are extremely heavy-tailed: in GPT-2
the ratio of the largest to the median production damage exceeds two orders of
magnitude. Any \emph{linear} standardization therefore leaves the two variances
mismatched (even a robust median-absolute-deviation scaling left a forty-fold ratio) and a difference of two mismatched-variance terms is
dominated by the larger. We built three such indices in turn, standardizing by
baseline across-item variability, by a robust scale estimated among
severity-matched peers, and by the robust across-site spread. Each one correlated
between $-0.88$ and $-0.97$ with the total size of the lesion, and between $+0.94$
and $+0.99$ with its own production term: all three were measuring how much the
ablation broke the model, and each flagged the largest lesions in every
architecture as ``selective''.

A rank transform removes the confound exactly rather than approximately. Ranking
each damage across sites makes both marginals uniform, hence their variances
equal, and since $\mathrm{cov}(x-y,\,x+y)=\mathrm{var}(x)-\mathrm{var}(y)$, the
selectivity contrast becomes exactly uncorrelated with total damage; measured, its
correlation with lesion size is $+0.01$ and with an independent severity measure
$-0.00$, with no nuisance regression at all. But the rank difference is bounded in
$[-1,+1]$, and a family-wise permutation null for a bounded statistic saturates at
its own bound: over several hundred sites the null maximum reaches $0.96$ against a
possible $1.00$, so no site can exceed it. Relaxing the null does not help, because
a per-site bootstrap over items tests whether a site's damage is non-zero rather
than whether the site is unusual among its peers, and it flags ordinary sites. The
per-site test is thus either valid without power or powered without validity.

\paragraph{Result 3: the two scores share a term, which manufactures coupling.}
A pooled statistic escapes the multiplicity problem, so we asked instead whether
comprehension damage and production damage co-vary \emph{across} sites at matched
lesion severity, using the divergence of the next-token distribution from its
unablated value as a read/write-neutral severity covariate. They do, in all three
models tested, with rank correlations of $+0.22$ to $+0.48$ against a stratified
permutation null. The result is nonetheless void, for a reason internal to the two
measures. Writing $a$, $b$ and $c$ for the mean log-probabilities of the target,
the competitor and the argmax token, the comprehension margin is $a-b$ and the
production capability is $a-c$: both contain $+a$. A site that moves the target's
own log-probability therefore moves both damages in the same direction with no
differential effect whatever. That is not a small correction: the movement in $a$
alone accounts for $96\%$ of the variance in production damage and $57\%$ in
comprehension damage. Removing the shared term algebraically ($b$ and $c$ are recoverable from what is logged, and share nothing) leaves a coupling of
$+0.13$, marginal against its permutation null, with two thirds of the association
gone; removing it by regression leaves nothing distinguishable from the null at
all. The measured coupling is, to a first approximation, a restatement of the fact
that both scores are anchored to the same log-probability.

\paragraph{An encoder/decoder proxy does not substitute.}
One might avoid output-side scores altogether by asking whether an architecture
with physically distinct stacks separates the two pathways representationally, and
comparing encoder and decoder hidden states on the same text is the obvious
measure. It does not work either, for three reasons we can state briefly. First,
the comparison is decided before any data is collected: a model with two disjoint
stacks must yield a large difference and a single-stack model must yield
approximately none, so the architecture class fixes the answer. Second, the measure
is insensitive to language. Shuffling the tokens within each sentence, which
destroys every syntactic and semantic relation while preserving the token
inventory, changes the encoder--decoder separation by $+0.02$ on average across the
four encoder--decoders we tested, and in two of them the shuffled text separates
\emph{more}. Whatever the number reflects, it is present for arbitrary token
sequences (consistent with its being the difference between a bidirectional
stack and a causal stack reading the same tokens shifted by one position). Third,
the estimate is unstable in the stimulus set: measured on hundreds of distinct
corpus sentences rather than a handful of repeated frames, two of the four models
fall by roughly a third, the ordering with model size stops being monotonic, and an
apparent robustness across training corpora disappears. We therefore make no
representational-separation claim for encoder--decoders.

\paragraph{What the three results amount to.}
None of this shows that reading and writing are indistinguishable inside these
models; it shows that they are not separable by the output-side operations
available to us, and it identifies why. Comprehension and production scores are
both functionals of one next-token distribution, so any two scalar summaries of it
share terms; the shared terms then generate the very coupling one is trying to
detect, and removing them removes the signal along with the artifact. This is a
sharper statement than the behavioural coupling we originally reported, and it is
the honest one: we can measure the entanglement of the reading and writing
\emph{codes} in the weights, where an independent-init floor and a tied ceiling
give the index a scale, but we cannot convert that into a behavioural double
dissociation test, and we now understand that as a consequence of the
single-forward-path design rather than a limitation of our stimuli. Whether a
richer probe (multi-token generation scored against a reference, or a model organism trained with deliberately separated read and write pathways) can recover a genuine dissociation test is open, and we regard it as the natural
next experiment rather than a gap in this one.

\section{Discussion}
\label{sec:discussion}

\subsection{What the measurements support, and what they cannot}

The positive result of this paper is a calibrated measurement at one level. In the
untied Pythia models the input embedding $\WE$ (the ``reading code'') and the output
unembedding $\WU$ (the ``writing code'') are \emph{coupled but sub-ceiling}: the
entanglement index settles roughly a third of the way up its range, far above the
independent-initialization floor yet well below the tied ceiling that GPT-2 and OPT
realize by construction. The number is interpretable because it has both a floor
and a ceiling; the trajectory
adds that the coupling is not monotone but a couple-then-differentiate program, and
that the differentiation is carried almost entirely by $\WU$.

We set out to pair this with a behavioural double dissociation test and could not
build one that was valid, across three designs (Section~\ref{sec:res-limits}). It
would be easy to present that as a limitation and move on. We think it is more than
that, because the three failures share a cause. Comprehension and production, as
anything at the output can define them, are both functionals of one next-token
distribution; any two scalar summaries of that distribution therefore share terms,
and the shared terms generate exactly the coupling one is trying to detect. In the
contingency design the shared structure is set-theoretic: greedy production
implies a vocabulary-wide argmax, so production's successes are a subset of
comprehension's, and for single-token pairs the alexia cell is empty by theorem
rather than by measurement. In the lesion design it is algebraic: the
comprehension margin and the production capability both contain the target's own
log-probability, whose movement alone accounts for most of the variance in each, so
their apparent co-variation is largely a restatement of that fact. Between these
sits a purely statistical obstacle: with damage distributions as heavy-tailed as
these, no linear standardization puts the two on a common scale, and the rank
transform that does is bounded, which costs the family-wise test its power.

The upshot is a sharper claim than the behavioural coupling we originally reported,
and it does not depend on any index we constructed. Reading and writing in a
decoder-only model are not merely coupled in the weights; at the output they are not
separately addressable at all. That is what one should expect of a single forward
path, and it is worth saying plainly rather than leaving as an unexamined
architectural assumption. It is also why we are careful, below, not to read the
weight-level coupling as though a behavioural dissociation had been ruled out
empirically. It has not been tested; we argue it cannot be, by these means.

\subsection{Reading without a reader}

This matters because it inverts the organization observed in the literate human
brain. There, reading and writing \emph{doubly dissociate}: pure alexia spares
writing while abolishing reading \citep{dejerine1892}, and pure agraphia spares
reading while abolishing writing \citep{caramazza1987buffer, roux2009exner}, with only a partial shared orthographic core
\citep{purcell2017shared, coltheart2001drc}. This $\WE$/$\WU$
contrast is an \emph{analogy, not a homology} (see Section~\ref{sec:limitations}): alexia and
agraphia motivate the question but are not measured here, and we intend no claim
of literal reading, writing, or ``understanding.'' With that disclaimer in force,
the contrast is informative precisely because the model and
the brain fall on opposite sides of it. The brain builds literacy by recruiting
\emph{two} pre-existing, dissociable circuits onto a culturally invented artifact
\citep{dehaene2007recycling,dehaene2011vwfa,morais1979}; the decoder-only LLM
builds the same functional competence by \emph{collapsing} decoding and encoding
into one autoregressive mechanism. Behaviourally it shows \emph{no} read/write
dissociation (Section~\ref{sec:results}); in this sense it ``reads without a
reader and writes without a writer.'' We are careful to scope this to behaviour:
we have not performed the model analog of a lesion study (a localized
intervention that selectively removes comprehension while sparing production, or
vice versa), so the stronger claim that \emph{no} dissociable component exists is
a conjecture we leave to future work, not a measured result.

This places decoder-only language models as a \emph{distinct point in the space
of possible minds} \citep{sloman1984,shanahan2025}: systems that attain a
recent cultural skill (written language, an invention of roughly five
millennia, not an evolved faculty; \citealp{pinker1994}) through an architecture
whose internal division of labour bears no structural resemblance to the brain
that invented the artifact. We read this not as a deficiency but as a
characterization. Following \citet{shanahan2025} and
\citet{mitchellkrakauer2023}, the appropriate move is to catalogue the mode of
intelligence on its own terms rather than to score it against a human template,
and our two-level convergence supplies one concrete, measurable axis (read/write
entanglement) on which the machine and the brain are organized oppositely.

\subsection{The development of the entangled code}

The training trajectory clarifies how the single code arises. In pythia-160m the
$\WE$/$\WU$ relationship is \emph{non-monotonic}: from an essentially
uncorrelated initialization the two matrices first couple sharply, peak early in
training, and then partially \emph{differentiate} for the remainder of it. The
differentiation is asymmetric, the writing code drifting several times farther
from initialization than the reading code, and by the same factor in every
frequency decile, which rules out the frequent-token gradient-exposure
explanation (Section~\ref{sec:res-entanglement}). As noted
above, we credit \citet{lopardo2026weighttying} for first reporting this
asymmetric output-side drift; our controlled, read/write-axis replication is
detailed in Section~\ref{sec:res-entanglement}. The picture is a
``couple-then-differentiate'' developmental program that nonetheless terminates
in coupling, not dissociation.

\subsection{Taking the nulls seriously}

We regard the negative results as load-bearing, not embarrassing, and report
them plainly. Three are central.

\paragraph{The geometry$\rightarrow$behaviour bridge is not merely null; it is
unavailable.} We had intended to ask whether the \emph{degree} of weight entanglement
predicts the \emph{degree} of behavioural coupling, and we reported that correlation as
a null. We now withdraw the test rather than the finding, because its behavioural term
was the degenerate coupling statistic of Section~\ref{sec:res-limits}: correlating
$\Eindex$ against a quantity that is an exact function of its own margins tests
nothing. Two obstacles would remain even with a sound behavioural measure. The tied
models sit degenerately at $\Eindex = 1$ with no variance and cannot enter a
cross-model test at all. And within training, production is undefined at the early
checkpoints where $\Eindex$ is low and changing, because the model cannot yet generate
the target forms; behaviour becomes measurable only \emph{after} the geometry has
plateaued, so the two quantities have little overlapping dynamic range. A cross-level
bridge is therefore future work contingent on a valid behavioural measure, and we
claim no mechanistic chain from geometry to behaviour.

\paragraph{Compression is not decoupled from functional competence.}
We had hypothesized that improvements in raw text compression might come apart
from improvements in functional, meaning-sensitive competence. We find no evidence
for that decoupling. Across the six Pythia sizes the point estimate runs in the
\emph{opposite} direction (bits-per-byte vs.\ functional accuracy Pearson
$r = -0.68$, $p = 0.14$; Spearman $\rho = -0.66$, $p = 0.16$; $n=6$): if anything
better compression accompanies better functional performance. With $n=6$ this is
underpowered and cannot establish co-variation either; the defensible conclusion
is simply that the ``compression decoupled from reasoning'' sub-claim is
\emph{not supported}, and we withdraw it as a positive claim.
This does not overturn the broader form-before-meaning pattern: formal
competence (BLiMP) leads functional competence across the ladder, with a gap of
a clear margin even at the largest model we test. The specific decoupling
mechanism we proposed, however, is absent here.

\paragraph{Mechanistic and scale links are weaker than hoped.}
Two further negatives constrain our claims. Ablating the top induction head
(L4H6) leaves the orthographic word-form gap intact (the gap after ablation is indistinguishable from the clean gap, and from a matched random-head control) so orthographic decoding is \emph{not} carried by the induction circuit, even
though both emerge in temporally aligned phases. The cross-model link between
maximal induction and in-context gain is positive but non-significant, and the $\Eindex$-versus-scale trend is flat and underpowered
(Spearman $\rho = 0.37$, permutation $p = 0.49$, $n=6$). We therefore scope the
induction claim to ``induction circuits are causally implicated in in-context
learning'' rather than to any architecture-wide or scale-wide generalization.

\section{Limitations}
\label{sec:limitations}

\paragraph{Analogy, not homology.}
As emphasized throughout, $\WE$ and $\WU$ are not reading and writing networks.
They are matrices that map tokens into and out of the residual stream
\citep{elhage2021framework}, operating in a single modality (discrete text) at a
single architectural stage. The brain's alexia/agraphia dissociation involves
distinct anatomical systems across distinct stages of perception and production
\citep{purcell2017shared}. Our comparison is a structured analogy that we believe
is illuminating, but every quantitative claim concerns the model alone; the
neuroscience is the motivating frame, not a co-measured variable.

\paragraph{The architecture nearly forces the result.}
A decoder-only transformer is architecturally committed to a single forward path,
so some degree of read/write coupling is expected a priori. The independent-init
floor shows that the matrices \emph{can} be fully dissociated at initialization, so the learned coupling is not a trivial
consequence of dimensionality; but the finding should be read as
``learning drives the codes toward a shared geometry that the architecture
permits and rewards,'' not as a discovery that two free systems chose to merge.

\paragraph{Tied-versus-untied is confounded across model families.}
The tied condition is realized only by the GPT-2 and OPT families and the untied
condition only by Pythia \citep{presswolf2017,inan2017tying}. Tying therefore
co-varies with training data, tokenizer, and optimization recipe. The tied
ceiling $\Eindex = 1.000$ is true by construction and is reported as a sanity
check, not as a finding; the substantive numbers come from the untied Pythia
suite \citep{biderman2023}, where the aligned checkpoints let us hold the recipe
fixed across scale and training time. A clean within-family tied-vs-untied
manipulation would strengthen the architectural claim and is left to future work.

\paragraph{Base models only; and no architecture contrast.}
All models studied are base pretrained LMs; instruction tuning and RLHF most directly
reshape \emph{production} behaviour, so claims about output-side behaviour are scoped
to base models. We had also intended to test whether read/write coupling is specific
to the decoder-only forward path, by contrasting decoder-only, encoder--decoder and
masked-encoder models behaviourally and by measuring how differently an
encoder--decoder's two stacks represent the same text. Both halves are withdrawn
(Section~\ref{sec:res-limits}): the behavioural half inherits the contingency defect,
and the representational half is fixed by architecture class before any data is
collected, is unchanged when linguistic structure is destroyed, and is unstable in the
stimulus set. We therefore make no cross-architecture claim in either direction. This
is a real gap: the weight-level entanglement is measured only on decoder-only suites,
so whether an architecture with physically separated pathways would show a different
read/write geometry is open. A matched-data, matched-tokenizer manipulation of
architecture alone, with a behavioural measure that survives the objections in
Section~\ref{sec:res-limits}, is the experiment we would run next.

\paragraph{Small samples and few-point fits.}
Several headline relationships rest on small $n$: the scale trend and the
compression--functional correlation use $n=6$, and the bridge tests use
$n \leq 12$ across models and $n=9$ within training. We report bootstrap
$95\%$ confidence intervals where available and label few-point fits as
\emph{suggestive}; the corresponding nulls are stated as underpowered rather than
as positive evidence of no effect.

\paragraph{Stimuli, tokenization, and the human comparison.}
The situation-tracking and absurdity items in the functional battery are
author-constructed and templated (frozen across runs for reproducibility), with
no human ceiling, so functional accuracy should be read as performance on
\emph{these} probes rather than on meaning in general. All measures inherit the
BPE tokenizer, which we control for with token-count and subword-unigram matching
but cannot fully neutralize. Finally, the comparison of machine token budgets to
the $\sim$100M-word human-by-age-13 reference \citep{warstadt2023babylm} is
order-of-magnitude and \emph{illustrative}: children receive multimodal,
embodied, interactive input, whereas these models see web text, so the contrast
indexes a gross efficiency gap, not a controlled equivalence
\citep{chollet2019measure}.

\section{Conclusion}
\label{sec:conclusion}

Written language is a recent cultural invention, not an evolved instinct
\citep{pinker1994,dehaene2007recycling}, and in the literate brain it is carried
by two doubly-dissociable systems for decoding and encoding that share only a
partial orthographic core. A decoder-only language model attains the same
functional skill by a structurally opposite route: it collapses reading and
writing into one entangled autoregressive code. We operationalized this contrast
through the reading code $\WE$ and the writing code $\WU$ and a composite
entanglement index $\Eindex$, and we found in the weights that the code is
single rather than dissociated: untied codes coupled well above the independent-init
floor yet well below the tied ceiling, on a couple-then-differentiate training
trajectory with an asymmetric output-side drift that replicates
\citet{lopardo2026weighttying}. We were equally explicit about the limits. The
matching behavioural test cannot be run: comprehension and production, as
output-side scores define them, are functionals of one next-token distribution and
share terms that manufacture the coupling they appear to reveal, so a coupling
statistic, a cross-level bridge and an encoder/decoder separation measure are
withdrawn. And the ``compression decoupled from reasoning'' hypothesis is not
supported, because compression and functional competence co-vary with scale.

The broader lesson speaks to the debate over the path to general intelligence.
These models demonstrate that a single general mechanism (self-attention;
\citealp{vaswani2017,olsson2022}) optimized on a cultural artifact can reach
impressive functional literacy without reconstructing any of the brain's
architecture for that skill, mastering form before meaning
\citep{mahowald2024,fedorenko2024language} and acting in part as a repository of
ingested text rather than a grounded abstraction
\citep{carlini2023quantifying,harnad1990symbol,yiu2024imitation,bender2021parrots}.
Rather than asking whether such a system is on the road to human-like
understanding, our results recommend treating it as what it measurably is: a
distinct point in the space of possible minds
\citep{sloman1984,shanahan2025,mitchellkrakauer2023}, one that reads and writes
through a single statistical code where the brain keeps two. Charting that space,
rather than collapsing it onto a human yardstick, is the lens we offer for the
next round of the question.

{\scriptsize
\setlength{\bibsep}{1.0pt}
\bibliographystyle{abbrvnat}
\bibliography{references}
}

\onecolumn  
\appendix
\renewcommand{\thefigure}{A\arabic{figure}}
\setcounter{figure}{0}

\section{Supplementary figures}
\label{sec:appendix}

The four figures collected here support claims established in the main text but
do not carry an independent argument: two are single-model illustrations (the
per-head induction map and the cross-model induction-vs-ICL scatter), and two
visualize underpowered null relationships (the entanglement index against scale,
and the geometry$\rightarrow$behaviour bridge). We place them here to keep the
main narrative focused while preserving the full evidentiary record.

\begin{figure}[h!]
  \centering
  \includegraphics[width=0.45\linewidth]{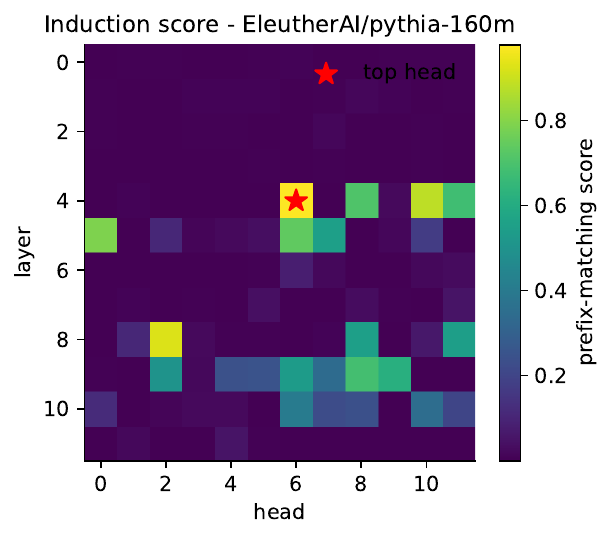}
  \caption{Per-head prefix-matching induction scores for pythia-160m. Induction
  behaviour is concentrated in a few heads rather than spread across the layer
  (the specialized-head pattern that justifies ablating a single top-scoring head
  in Section~\ref{sec:res-mechanism}; \citealp{voita2019, olsson2022}).}
  \label{fig:exp1_heatmap}
\end{figure}

\begin{figure}[h!]
  \centering
  \includegraphics[width=0.65\linewidth]{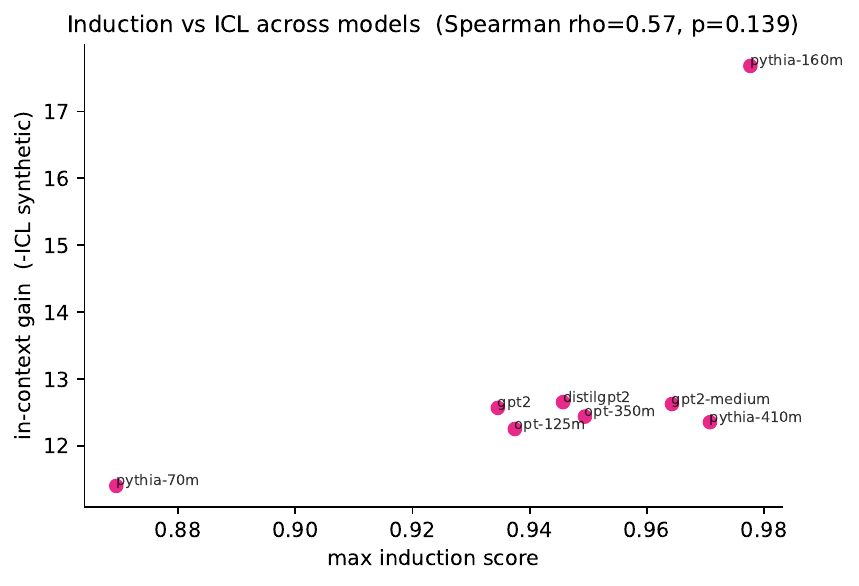}
  \caption{Maximum induction score against in-context gain across the eight
  models. The association is positive but non-significant (Spearman $\rho=0.57$,
  $p=0.139$, $n=8$); this small-$n$ cross-model trend is why the causal claim in
  the main text rests on within-model ablation rather than on this correlation.}
  \label{fig:exp1_scatter}
\end{figure}

\begin{figure}[h!]
  \centering
  \includegraphics[width=0.6\linewidth]{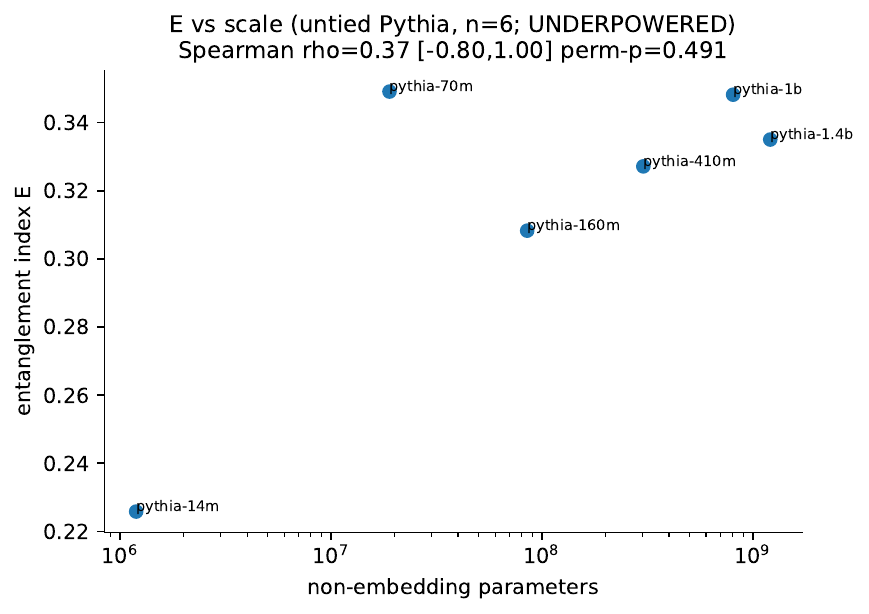}
  \caption{Entanglement index $\Eindex$ against model scale for the six untied
  Pythia models. There is no systematic trend (Spearman $0.37$, perm-$p=0.49$,
  $n=6$): the degree of read/write coupling is approximately scale-invariant over
  this range, though $n=6$ is too small to assert flatness rather than merely
  fail to detect a slope.}
  \label{fig:exp7_scale}
\end{figure}

\renewcommand{\thetable}{B\arabic{table}}
\setcounter{table}{0}

\section{Supplementary tables}
\label{sec:appendix-tables}

The tables below collect every numerical result referenced in the main text, in
order of first mention and renumbered B1--B11. Table~\ref{tab:exp3_data_dynamics}
gives the full tokens-to-competence trajectory across all released checkpoints.

\begin{table}
\centering
\caption{Induction-head localization and causal ablation across eight small open models. $\Delta$ICL is the change in the synthetic copy-completion in-context score after zero-ablating the top induction head vs the MLP block at the same layer; head $z$ is the top head's $\Delta$ICL relative to a 24-random-head null. Cross-model Spearman(max induction, in-context gain)$=0.57$ ($p=0.139$).}
\label{tab:exp1_induction}
\adjustbox{max width=\linewidth}{%
\begin{tabular}{lrrrrrrr}
\toprule
Model & L & H & Max induction [95\% CI] & ICL (syn.) & $\Delta$ICL ablate head & $\Delta$ICL ablate MLP & head $z$ vs null \\
\midrule
distilgpt2 & 6 & 12 & 0.946 [0.944, 0.947] & -12.656 & 0.097 & 0.258 & -0.091 \\
gpt2 & 12 & 12 & 0.935 [0.933, 0.937] & -12.570 & -0.056 & 0.123 & -1.502 \\
gpt2-medium & 24 & 16 & 0.964 [0.963, 0.965] & -12.627 & 0.009 & 0.109 & 0.077 \\
facebook/opt-125m & 12 & 12 & 0.937 [0.937, 0.938] & -12.254 & 0.466 & 0.144 & 9.572 \\
facebook/opt-350m & 24 & 16 & 0.949 [0.947, 0.951] & -12.440 & 0.039 & 0.096 & 0.477 \\
EleutherAI/pythia-70m & 6 & 8 & 0.869 [0.866, 0.873] & -11.404 & 1.413 & 2.158 & 0.885 \\
EleutherAI/pythia-160m & 12 & 12 & 0.978 [0.977, 0.979] & -17.680 & 1.652 & 0.650 & 4.249 \\
EleutherAI/pythia-410m & 24 & 16 & 0.971 [0.970, 0.972] & -12.356 & 0.094 & -0.270 & 2.069 \\
\bottomrule
\end{tabular}}

\end{table}

\begin{table}
\centering
\caption{Emergence of orthographic word-form selectivity over Pythia training. Word-form gap $=$ bpb(illegal)$-$bpb(real), scored over word-form continuation tokens only (carrier masked). The raw gap is already large at initialization and \emph{shrinks} with training (a BPE tokenization artifact, since illegal strings fragment into more sub-tokens), whereas the genuine acquired signal (the token-count-controlled gap, both residualized and length-matched) \emph{grows} with training. Onset (max-slope) step-gap word-form$-$ICL $=-255$, word-form$-$induction $=-999$; CI overlap of onsets is NOT evidence of a shared mechanism. Step B-causal: ablating the top induction head leaves $1.01$ of the word-form gap (random-head control delta $=-0.001$), so gap collapse $=$ False.}
\label{tab:exp2_wordform_dynamics}
\adjustbox{max width=\linewidth}{%
\begin{tabular}{lrrrrrr}
\toprule
Model & Word-form gap @init & Word-form gap @final & Token-ctrl gap @final & Tok-matched gap @final & Legality gap @final & Legality tok-ctrl @final \\
\midrule
pythia-160m & 5.883 [5.839, 5.922] & 4.375 [4.325, 4.436] & 1.854 [1.739, 1.952] & 3.795 [3.583, 4.007] & 1.652 [1.617, 1.706] & 0.594 [0.585, 0.608] \\
pythia-410m & 5.891 [5.825, 5.954] & 4.535 [4.493, 4.565] & 2.034 [1.880, 2.221] & 4.366 [4.204, 4.518] & 1.816 [1.762, 1.857] & 0.767 [0.738, 0.817] \\
\bottomrule
\end{tabular}}

\end{table}

\begin{table}[t]
\centering
\caption{Linear-probe decodability of orthographic word-form class (real/legal/illegal) per layer. The real-vs-scrambled-real control (token-count matched) tests whether decodability reflects orthographic structure rather than token-count leakage. Permutation null shown for the best layer.}
\label{tab:exp2_probe}
\adjustbox{max width=\linewidth}{%
\begin{tabular}{lrrrrr}
\toprule
Model & Best layer & 3-class acc & Chance & Perm null & Real-vs-scrambled (best layer) \\
\midrule
EleutherAI/pythia-160m@step143000 & 8 & 0.933 & 0.333 & 0.352 & 0.869 \\
gpt2 & 9 & 0.935 & 0.333 & 0.331 & 0.828 \\
\bottomrule
\end{tabular}}

\end{table}

{\scriptsize
\setlength{\tabcolsep}{4pt}
\begin{longtable}{lrrrrr}
\caption{Tokens-to-competence on released Pythia checkpoints (no training). The perplexity point estimate is corpus-level (token-weighted, $\exp(\sum\mathrm{nll}/\sum\mathrm{tok})$) while the bracketed range is a per-document bootstrap of the unweighted mean log-perplexity; these are different estimands, so the point estimate need not fall inside the range. Machine budget shown in tokens AND approximate word-equivalents (tokens$/1.30$). The $\sim$100M-word human-by-age-13 figure is an ILLUSTRATIVE order-of-magnitude reference, not a cognitive equivalence (humans get multimodal/embodied/interactive child-directed input; wikitext/Pile is adult written text); the BabyLM budget (warstadt2023babylm) is a budget target only. Across-size loss-vs-$N$ power law at step143000: $\alpha_N=0.100$ [-0.045, 0.113] (no offset), $0.000$ (with offset). WITHIN-RUN fits are autocorrelated (CI understates uncertainty). BLiMP chance$=0.5$; step1--4 is the near-chance floor.}
\label{tab:exp3_data_dynamics}\\
\toprule
Model & Ckpt & Tokens (word-eq.) & Loss (nats) & Perplexity (doc-level range) & BLiMP acc [95\% CI] \\
\midrule
\endfirsthead
\multicolumn{6}{l}{\footnotesize\itshape Table \ref{tab:exp3_data_dynamics}, continued}\\
\toprule
Model & Ckpt & Tokens (word-eq.) & Loss (nats) & Perplexity (doc-level range) & BLiMP acc [95\% CI] \\
\midrule
\endhead
\midrule
\multicolumn{6}{r}{\footnotesize\itshape continued on next page}\\
\endfoot
\bottomrule
\endlastfoot
EleutherAI/pythia-160m & step1 & 2.10e+06 (1.61e+06) & 11.044 & 62554.8 [61964.9, 63169.2] & 0.484 [0.463, 0.507] \\
EleutherAI/pythia-160m & step2 & 4.19e+06 (3.23e+06) & 11.043 & 62499.8 [61911.3, 63114.6] & 0.484 [0.463, 0.507] \\
EleutherAI/pythia-160m & step4 & 8.39e+06 (6.45e+06) & 11.005 & 60180.1 [59673.1, 60812.4] & 0.483 [0.461, 0.505] \\
EleutherAI/pythia-160m & step8 & 1.68e+07 (1.29e+07) & 10.844 & 51211.4 [50889.6, 52043.3] & 0.486 [0.465, 0.509] \\
EleutherAI/pythia-160m & step16 & 3.36e+07 (2.58e+07) & 10.473 & 35356.4 [34893.7, 36390.6] & 0.500 [0.478, 0.521] \\
EleutherAI/pythia-160m & step32 & 6.71e+07 (5.16e+07) & 9.992 & 21851.0 [21254.5, 22754.5] & 0.520 [0.498, 0.542] \\
EleutherAI/pythia-160m & step64 & 1.34e+08 (1.03e+08) & 9.321 & 11171.6 [10741.7, 11673.3] & 0.442 [0.420, 0.463] \\
EleutherAI/pythia-160m & step128 & 2.68e+08 (2.06e+08) & 8.186 & 3592.1 [3416.2, 3789.4] & 0.508 [0.487, 0.530] \\
EleutherAI/pythia-160m & step256 & 5.37e+08 (4.13e+08) & 7.430 & 1686.5 [1589.3, 1791.8] & 0.599 [0.577, 0.620] \\
EleutherAI/pythia-160m & step512 & 1.07e+09 (8.26e+08) & 6.565 & 709.9 [671.7, 764.6] & 0.587 [0.565, 0.608] \\
EleutherAI/pythia-160m & step1000 & 2.10e+09 (1.61e+09) & 5.432 & 228.5 [231.5, 270.9] & 0.647 [0.626, 0.668] \\
EleutherAI/pythia-160m & step2000 & 4.19e+09 (3.23e+09) & 4.707 & 110.7 [111.9, 130.0] & 0.716 [0.696, 0.736] \\
EleutherAI/pythia-160m & step4000 & 8.39e+09 (6.45e+09) & 4.323 & 75.4 [76.0, 87.7] & 0.756 [0.737, 0.774] \\
EleutherAI/pythia-160m & step8000 & 1.68e+10 (1.29e+10) & 4.121 & 61.6 [62.5, 71.2] & 0.753 [0.733, 0.772] \\
EleutherAI/pythia-160m & step16000 & 3.36e+10 (2.58e+10) & 3.984 & 53.7 [54.5, 62.6] & 0.772 [0.752, 0.790] \\
EleutherAI/pythia-160m & step32000 & 6.71e+10 (5.16e+10) & 3.905 & 49.7 [50.3, 57.5] & 0.757 [0.738, 0.776] \\
EleutherAI/pythia-160m & step64000 & 1.34e+11 (1.03e+11) & 3.861 & 47.5 [48.0, 54.7] & 0.762 [0.742, 0.780] \\
EleutherAI/pythia-160m & step143000 & 3.00e+11 (2.31e+11) & 3.904 & 49.6 [49.7, 56.5] & 0.744 [0.725, 0.763] \\
EleutherAI/pythia-410m & step1 & 2.10e+06 (1.61e+06) & 11.058 & 63420.1 [63017.6, 64207.2] & 0.551 [0.528, 0.572] \\
EleutherAI/pythia-410m & step2 & 4.19e+06 (3.23e+06) & 11.055 & 63272.2 [62870.7, 64067.1] & 0.551 [0.529, 0.573] \\
EleutherAI/pythia-410m & step4 & 8.39e+06 (6.45e+06) & 10.965 & 57816.1 [57508.4, 58802.2] & 0.564 [0.542, 0.585] \\
EleutherAI/pythia-410m & step8 & 1.68e+07 (1.29e+07) & 10.686 & 43722.2 [43343.7, 45066.2] & 0.583 [0.561, 0.605] \\
EleutherAI/pythia-410m & step16 & 3.36e+07 (2.58e+07) & 10.242 & 28063.6 [27459.6, 29307.7] & 0.538 [0.516, 0.560] \\
EleutherAI/pythia-410m & step32 & 6.71e+07 (5.16e+07) & 9.930 & 20529.2 [19859.2, 21515.3] & 0.512 [0.490, 0.534] \\
EleutherAI/pythia-410m & step64 & 1.34e+08 (1.03e+08) & 9.487 & 13189.4 [12623.7, 13878.4] & 0.475 [0.453, 0.497] \\
EleutherAI/pythia-410m & step128 & 2.68e+08 (2.06e+08) & 8.615 & 5515.5 [5283.7, 5847.2] & 0.497 [0.476, 0.519] \\
EleutherAI/pythia-410m & step256 & 5.37e+08 (4.13e+08) & 7.660 & 2122.0 [2010.7, 2263.1] & 0.558 [0.536, 0.579] \\
EleutherAI/pythia-410m & step512 & 1.07e+09 (8.26e+08) & 6.897 & 989.3 [928.9, 1055.5] & 0.583 [0.561, 0.605] \\
EleutherAI/pythia-410m & step1000 & 2.10e+09 (1.61e+09) & 5.497 & 244.1 [247.8, 287.2] & 0.660 [0.639, 0.680] \\
EleutherAI/pythia-410m & step2000 & 4.19e+09 (3.23e+09) & 4.513 & 91.2 [91.6, 105.8] & 0.696 [0.675, 0.716] \\
EleutherAI/pythia-410m & step4000 & 8.39e+09 (6.45e+09) & 4.078 & 59.0 [59.9, 68.8] & 0.758 [0.738, 0.777] \\
EleutherAI/pythia-410m & step8000 & 1.68e+10 (1.29e+10) & 3.842 & 46.6 [47.6, 54.3] & 0.782 [0.764, 0.801] \\
EleutherAI/pythia-410m & step16000 & 3.36e+10 (2.58e+10) & 3.645 & 38.3 [38.8, 44.3] & 0.806 [0.788, 0.824] \\
EleutherAI/pythia-410m & step32000 & 6.71e+10 (5.16e+10) & 3.557 & 35.1 [35.6, 40.5] & 0.781 [0.763, 0.800] \\
EleutherAI/pythia-410m & step64000 & 1.34e+11 (1.03e+11) & 3.475 & 32.3 [32.8, 37.5] & 0.777 [0.758, 0.795] \\
EleutherAI/pythia-410m & step143000 & 3.00e+11 (2.31e+11) & 3.381 & 29.4 [29.8, 34.2] & 0.805 [0.787, 0.823] \\
EleutherAI/pythia-70m & step1 & 2.10e+06 (1.61e+06) & 11.002 & 59996.7 [59729.3, 60941.4] & 0.491 [0.469, 0.513] \\
EleutherAI/pythia-70m & step2 & 4.19e+06 (3.23e+06) & 11.002 & 59982.0 [59714.8, 60926.2] & 0.491 [0.470, 0.513] \\
EleutherAI/pythia-70m & step4 & 8.39e+06 (6.45e+06) & 10.997 & 59710.0 [59444.3, 60651.9] & 0.493 [0.472, 0.515] \\
EleutherAI/pythia-70m & step8 & 1.68e+07 (1.29e+07) & 10.937 & 56202.6 [55982.7, 57129.1] & 0.495 [0.474, 0.517] \\
EleutherAI/pythia-70m & step16 & 3.36e+07 (2.58e+07) & 10.745 & 46409.4 [46217.0, 47296.7] & 0.501 [0.479, 0.523] \\
EleutherAI/pythia-70m & step32 & 6.71e+07 (5.16e+07) & 10.325 & 30483.4 [30025.6, 31435.9] & 0.525 [0.503, 0.546] \\
EleutherAI/pythia-70m & step64 & 1.34e+08 (1.03e+08) & 9.587 & 14572.1 [14210.9, 15168.6] & 0.584 [0.563, 0.606] \\
EleutherAI/pythia-70m & step128 & 2.68e+08 (2.06e+08) & 8.172 & 3539.5 [3392.0, 3735.9] & 0.514 [0.492, 0.537] \\
EleutherAI/pythia-70m & step256 & 5.37e+08 (4.13e+08) & 7.408 & 1649.8 [1543.5, 1747.6] & 0.587 [0.566, 0.609] \\
EleutherAI/pythia-70m & step512 & 1.07e+09 (8.26e+08) & 6.596 & 732.5 [682.5, 777.7] & 0.572 [0.551, 0.594] \\
EleutherAI/pythia-70m & step1000 & 2.10e+09 (1.61e+09) & 5.531 & 252.3 [254.2, 298.3] & 0.609 [0.587, 0.631] \\
EleutherAI/pythia-70m & step2000 & 4.19e+09 (3.23e+09) & 4.969 & 143.9 [144.2, 168.2] & 0.682 [0.661, 0.702] \\
EleutherAI/pythia-70m & step4000 & 8.39e+09 (6.45e+09) & 4.728 & 113.1 [114.2, 132.5] & 0.720 [0.699, 0.739] \\
EleutherAI/pythia-70m & step8000 & 1.68e+10 (1.29e+10) & 4.595 & 99.0 [100.2, 116.5] & 0.739 [0.720, 0.758] \\
EleutherAI/pythia-70m & step16000 & 3.36e+10 (2.58e+10) & 4.497 & 89.7 [90.3, 104.5] & 0.713 [0.692, 0.732] \\
EleutherAI/pythia-70m & step32000 & 6.71e+10 (5.16e+10) & 4.444 & 85.1 [85.3, 98.2] & 0.718 [0.698, 0.737] \\
EleutherAI/pythia-70m & step64000 & 1.34e+11 (1.03e+11) & 4.433 & 84.2 [85.2, 98.1] & 0.720 [0.699, 0.739] \\
EleutherAI/pythia-70m & step143000 & 3.00e+11 (2.31e+11) & 4.464 & 86.8 [87.1, 100.1] & 0.718 [0.698, 0.737] \\
\end{longtable}
}

\begin{table}
\centering
\caption{Form before meaning. Formal competence (BLiMP grammaticality) is high and near-saturated across the Pythia scale ladder, while functional competence (PIQA physical commonsense, situation tracking, absurdity detection) rises more slowly and remains well below the formal level. All values are accuracy [bootstrap 95\% CI] under per-token mean log-prob scoring; chance is 0.5. The grounding gap is surface-control minus situation-tracking accuracy. No human ceiling exists for the custom situation set; pythia-1.4b anchors a reference upper bound.}
\label{tab:exp4_form_function}
\adjustbox{max width=\linewidth}{%
\begin{tabular}{lrrrrrr}
\toprule
Model & N (non-emb.) & BLiMP (formal) & PIQA & Situation & Absurdity & Grounding gap \\
\midrule
pythia-14m & 1.2M & 0.695 [0.674, 0.715] & 0.596 [0.552, 0.636] & 0.525 [0.433, 0.617] & 1.000 [1.000, 1.000] & -0.050 \\
pythia-70m & 18.9M & 0.718 [0.698, 0.737] & 0.576 [0.532, 0.618] & 0.333 [0.250, 0.417] & 1.000 [1.000, 1.000] & 0.500 \\
pythia-160m & 85.1M & 0.744 [0.725, 0.763] & 0.616 [0.570, 0.656] & 0.425 [0.333, 0.508] & 1.000 [1.000, 1.000] & 0.308 \\
pythia-410m & 302.3M & 0.805 [0.787, 0.823] & 0.670 [0.626, 0.708] & 0.717 [0.633, 0.800] & 1.000 [1.000, 1.000] & -0.225 \\
pythia-1b & 805.7M & 0.806 [0.789, 0.824] & 0.692 [0.650, 0.730] & 0.725 [0.642, 0.800] & 1.000 [1.000, 1.000] & -0.433 \\
pythia-1.4b & 1208.6M & 0.824 [0.807, 0.841] & 0.688 [0.646, 0.726] & 0.583 [0.492, 0.675] & 1.000 [1.000, 1.000] & 0.008 \\
\bottomrule
\end{tabular}}

\end{table}

\begin{table}[t]
\centering
\caption{Compression of held-out wikitext-2 as a function of non-embedding parameter count $N$. Power-law fit (no offset) $\alpha=0.090$ [0.080, 0.107], $R^2=0.984$; with offset $\alpha=0.000$.}
\label{tab:exp5_scaling}
\adjustbox{max width=\textwidth}{%
\begin{tabular}{lrrrr}
\toprule
Model & $N$ & CE (nats) & PPL & bits/byte \\
\midrule
EleutherAI/pythia-14m & 1189888 & 4.728 & 113.028 & 1.489 \\
EleutherAI/pythia-70m & 18915328 & 3.959 & 52.383 & 1.247 \\
EleutherAI/pythia-160m & 85056000 & 3.395 & 29.821 & 1.069 \\
EleutherAI/pythia-410m & 302311424 & 2.896 & 18.107 & 0.912 \\
EleutherAI/pythia-1b & 805736448 & 2.705 & 14.949 & 0.852 \\
EleutherAI/pythia-1.4b & 1208602624 & 2.593 & 13.364 & 0.816 \\
\bottomrule
\end{tabular}}

\end{table}

\begin{table}[t]
\centering
\caption{Neural vs Kneser-Ney BPE $n$-gram (order 3) on held-out wikitext-2. The interpretable signal is the SCALING of the log-PPL gap (nats/token the neural model saves over the $n$-gram), not the raw ratio. Top-1 next-token agreement with pythia-160m $=0.290$, symmetric KL $=3.030$ nats.}
\label{tab:exp5_ngram}
\adjustbox{max width=\textwidth}{%
\begin{tabular}{lrrrrr}
\toprule
Model & $N$ & neural PPL & n-gram PPL & ratio & log-PPL gap \\
\midrule
EleutherAI/pythia-14m & 1189888 & 104.467 & 473.479 & 0.221 & 1.511 \\
EleutherAI/pythia-70m & 18915328 & 48.577 & 473.479 & 0.103 & 2.277 \\
EleutherAI/pythia-160m & 85056000 & 27.922 & 473.479 & 0.059 & 2.831 \\
EleutherAI/pythia-410m & 302311424 & 16.234 & 473.479 & 0.034 & 3.373 \\
EleutherAI/pythia-1b & 805736448 & 13.383 & 473.479 & 0.028 & 3.566 \\
EleutherAI/pythia-1.4b & 1208602624 & 11.843 & 473.479 & 0.025 & 3.688 \\
\bottomrule
\end{tabular}}

\end{table}

\begin{table}[t]
\centering
\caption{EXP6 headline findings, reported SEPARATELY (not merged). P1: log-linear slope of verbatim Pile extraction vs log non-embedding params and vs log prefix length, contrasted with a non-member control. P2a: imitation$-$innovation accuracy gap vs a PPMI rarity baseline. P2b: best-layer |Spearman| between hidden-state geometry and perceptual order vs a swept PPMI baseline. The `repository-not-abstraction' synthesis is a hypothesis, not proven by any single probe.}
\label{tab:exp6_repository}
\adjustbox{max width=\textwidth}{%
\begin{tabular}{lrr}
\toprule
Finding & Headline & Baseline/control \\
\midrule
P1 memorization vs log(params) & 0.011 [0.007, 0.017] & non-member extraction (generic fluency) \\
P1 memorization vs log(prefix len) & 0.017 [-0.012, 0.045] & non-member extraction (generic fluency) \\
P2a imitation$-$innovation gap (max) & 0.090 & PPMI rarity gap 0.000 \\
P2b geometry vs perceptual |rho| (best) & 0.826 & PPMI |rho| max 0.512 \\
\bottomrule
\end{tabular}}

\end{table}

\begin{table}
\centering
\scriptsize
\caption{Read-code ($W_E$) vs write-code ($W_U$) entanglement. Tied models (GPT-2, OPT) return $E\approx1$ BY CONSTRUCTION ($W_E=W_U$; a pipeline sanity check, not a finding). Untied Pythia models sit high but measurably above the independent-init CKA floor: a single entangled code, not two dissociated systems. CIs are bootstrap 95\%. $W_E/W_U$ are input- vs output-side orthographic coding; the alexia/agraphia parallel is analogy, not homology.}
\label{tab:exp7_entanglement}
\adjustbox{max width=\textwidth}{%
\begin{tabular}{lrrrrrrr}
\toprule
Model & Tying & E [95\% CI] & CKA & Procrustes resid. & mean cos. & kNN@10 & CKA floor \\
\midrule
pythia-14m & untied & 0.226 [0.223, 0.228] & 0.182 & 1.142 & 0.401 & 0.159 & 0.003 \\
pythia-70m & untied & 0.349 [0.338, 0.360] & 0.353 & 1.015 & 0.475 & 0.396 & 0.010 \\
pythia-160m & untied & 0.308 [0.301, 0.315] & 0.210 & 0.966 & 0.454 & 0.451 & 0.015 \\
pythia-410m & untied & 0.327 [0.324, 0.330] & 0.295 & 1.124 & 0.404 & 0.534 & 0.020 \\
pythia-1b & untied & 0.348 [0.345, 0.351] & 0.339 & 1.105 & 0.439 & 0.585 & 0.039 \\
pythia-1.4b & untied & 0.335 [0.332, 0.338] & 0.317 & 1.109 & 0.421 & 0.588 & 0.039 \\
gpt2 & tied & 1.000 [1.000, 1.000] & 1.000 & 0.000 & 1.000 & 1.000 & 0.015 \\
gpt2-medium & tied & 1.000 [1.000, 1.000] & 1.000 & 0.000 & 1.000 & 1.000 & 0.020 \\
opt-125m & tied & 1.000 [1.000, 1.000] & 1.000 & 0.000 & 1.000 & 1.000 & 0.015 \\
opt-350m & tied & 1.000 [1.000, 1.000] & 1.000 & 0.000 & 1.000 & 1.000 & 0.010 \\
\bottomrule
\end{tabular}}

\end{table}

\begin{table}[t]
\centering
\caption{Per-token read/write correspondence against its null and floor (untied models). Observed mean per-token cosine and mutual-kNN overlap between $W_E$ and $W_U$ greatly exceed both the shuffled-token null (row-permuted $W_U$, destroying per-token correspondence) and the independent-initialization floor, confirming the coupling is genuine per-token structure, not a global-subspace artifact.}
\label{tab:exp7_nulls}
\adjustbox{max width=\textwidth}{%
\begin{tabular}{lrrrrrr}
\toprule
Model & cos (obs.) & cos (shuffle) & cos (floor) & kNN (obs.) & kNN (shuffle) & kNN (floor) \\
\midrule
pythia-14m & 0.401 & 0.038 & 0.043 & 0.159 & 0.000 & 0.000 \\
pythia-70m & 0.475 & 0.076 & 0.085 & 0.396 & 0.000 & 0.000 \\
pythia-160m & 0.454 & 0.094 & 0.104 & 0.451 & 0.000 & 0.000 \\
pythia-410m & 0.404 & 0.114 & 0.121 & 0.534 & 0.000 & 0.000 \\
pythia-1b & 0.439 & 0.159 & 0.171 & 0.585 & 0.000 & 0.000 \\
pythia-1.4b & 0.421 & 0.158 & 0.171 & 0.588 & 0.000 & 0.000 \\
\bottomrule
\end{tabular}}
\end{table}

\begin{table}[t]
\centering
\caption{Asymmetric read/write drift decomposed by token-frequency decile (pythia-160m, step143000). The output code $W_U$ drifts substantially more than the input code $W_E$ in every frequency decile, so the asymmetry is not an artifact of frequent tokens receiving more updates; it is the control that distinguishes this analysis from the single-trajectory report of \citet{lopardo2026weighttying}.}
\label{tab:exp7_decile_drift}
\adjustbox{max width=\textwidth}{%
\begin{tabular}{lrrr}
\toprule
Frequency decile & drift $W_E$ & drift $W_U$ & ratio $U/E$ \\
\midrule
0 (rarest) & 1.551 & 5.048 & 3.254 \\
1 () & 1.581 & 5.066 & 3.204 \\
2 () & 1.583 & 5.069 & 3.203 \\
3 () & 1.583 & 5.070 & 3.204 \\
4 () & 1.584 & 5.098 & 3.218 \\
5 () & 1.591 & 5.145 & 3.235 \\
6 () & 1.593 & 5.156 & 3.236 \\
7 () & 1.592 & 5.172 & 3.249 \\
8 () & 1.584 & 5.179 & 3.270 \\
9 (most freq.) & 1.533 & 5.180 & 3.379 \\
\bottomrule
\end{tabular}}
\end{table}

\end{document}